\documentclass[
reprint,
superscriptaddress,
showpacs,preprintnumbers,
amsmath,amssymb,
aps,
]{revtex4-1}

\usepackage{color}
\usepackage{graphicx}
\usepackage{dcolumn}
\usepackage{bm}
\usepackage{txfonts}
\usepackage{tikz}
\usepackage{pgffor}
\usepackage{verbatim}
\usepackage{pstricks}
\usepackage{pst-3dplot}
\usepackage[colorlinks, breaklinks=true, linkcolor=blue, citecolor=blue, linktocpage=true]{hyperref}

\begin{document}


\title{Circuit-based cavity magnonics in the ultrastrong and deep-strong coupling regimes}

\author{Takahiro Chiba}
 \affiliation{Department of Information Science and Technology, Graduate School of Science and Engineering, Yamagata University, Yonezawa, Yamagata 992-8510, Japan}
 \affiliation{Department of Applied Physics, Graduate School of Engineering, Tohoku University, Sendai, Miyagi 980-8579, Japan}
\author{Ryunosuke Suzuki}
 \affiliation{Department of Applied Physics, Graduate School of Engineering, Tohoku University, Sendai, Miyagi 980-8579, Japan}
\author{Takashi Otaki}
 \affiliation{Department of Applied Physics, Graduate School of Engineering, Tohoku University, Sendai, Miyagi 980-8579, Japan}
\author{Hiroaki Matsueda}
 \affiliation{Department of Applied Physics, Graduate School of Engineering, Tohoku University, Sendai, Miyagi 980-8579, Japan}
 \affiliation{Center for Science and Innovation in Spintronics, Tohoku University, Sendai 980-8577, Japan}


\date{\today}

\begin{abstract}
We theoretically study nonperturbative strong-coupling phenomena in cavity magnonics systems in which the uniform magnetization dynamics (magnons) in a ferromagnet is coupled to the microwave magnetic field (photons) of a single $LC$ resonator. 
Starting from an effective circuit model that accounts for the magnetization dynamics described by the Landau-Lifshitz-Gilbert equation, we show that a nontrivial frequency shift emerges in the ultrastrong and deep-strong coupling regimes, whose microscopic origin remains elusive within a purely classical framework.
The circuit model is further quantized to derive a minimal quantum mechanical model for generic cavity magnonics, which corresponds to a two-mode version of the Hopfield Hamiltonian and explains the mechanism of the frequency shifts found in the {\it classical} circuit model. 
We also formulate the relation between the frequency shift and quantum quantities, such as the ground-state particle number, quantum fluctuations associated with the Heisenberg uncertainty principle, and entanglement entropy, providing a nondestructive means to experimentally access to these quantum resources. 
By utilizing soft magnons in an anisotropic ferromagnet, we further demonstrate that these quantum quantities diverge at the zeros of the magnon band edges as a function of the external magnetic field. This work paves the way for cavity magnonics beyond the conventional strong coupling regime.

\end{abstract}

\maketitle

\section{Introduction} 

Cavity quantum electrodynamics (QED) constitutes a minimal framework for the study of light-matter interactions between an atomic ensemble and a single-mode electromagnetic wave (cavity-photon) at the quantum level \cite{Walther06}. A central focus of this field has been the realization and characterization of strong coupling, wherein the coherent interaction strength $g$ exceeds the dissipation rates of the system, leading to the formation of hybrid light-matter quasiparticles known as polaritons. Nonetheless, such strongly coupled states are still {\it perturbative} in nature, given that the interaction is considerably smaller than the atomic or photonic energy scales ($\omega$), i.e., $g\ll\omega$ (in units of angular frequency). Recent developments in both cavity QED and circuit QED platforms \cite{Blais21} have enabled access to regimes beyond this perturbative limit, facilitating the exploration of {\it nonperturbative} strong-coupling phenomena. These include the so-called ultrastrong coupling (USC, defined by $g/\omega \gtrsim 0.1$) and deep-strong coupling (DSC, $g/\omega \gtrsim 1$) regimes \cite{Diaz19,Kockum19,Schlawin22,Baydin25}. These regimes give rise to a variety of fundamentally novel and nontrivial effects, such as the vacuum Bloch-Siegert (BS) shift \cite{Diaz10,Li18}, non-zero ground-state photon populations (virtual photons) \cite{Hirokawa17}, an entangled cat-state between qubits and photons \cite{Yoshihara18}, superradiant phase transitions \cite{Hepp73,Bernardis18}, perfect intrinsic squeezing \cite{Hayashida23}, a quantum battery \cite{Ferraro18}, and more. Unfortunately, a few of these phenomena are experimentally verified so far \cite{Diaz10,Li18,Yoshihara18} and therefore it is highly desired to take the rest of unrevealed phenomena into experimentally testable stages.

Magnons---the elementary excitations of the ordered magnet---have garnered significant attention as potential information carriers for future computation and communication technologies owing to their advantageous features, including the ability to tune their eigenfrequency by an external magnetic field, a high spin density, stability at room temperature, and the absence of Joule heating \cite{Chumak22}. Combining magnon and cavity QED, the research field of ``cavity magnonics'' has emerged, which studies the strong coupling between magnons and cavity photons or the equivalent concept being magnon-polariton (MP) \cite{Yuan22,Rameshti22,Harder21}. Cavity magnonics have presented promising potential applications for quantum information technology, such as squeezed magnon states \cite{Elyasi20,Lee23,Yang21,Kani25}, quantum-enhanced metrology \cite{Wan24}, the macroscopic quantum state \cite{Sun21,Kounalakis22}, the entanglement generation \cite{Elyasi20,Mousolou21,Silaev23}, coherent microwave emission (maser-like behavior) due to non-Hermiticity \cite{Yao23,Zhang25}, and the enhancement of the magnon spin angular momentum beyond the standard value of $\hbar$ \cite{Kamra16} within the strong coupling regime.
More recently, several experiments have demonstrated the USC of MP by employing superconducting resonators \cite{Golovchanskiy21SciAdv,Golovchanskiy21PRAP,Ghirri23} at cryogenic temperature, as well as slab geometries \cite{Bourcin23} and magnetic metamaterials \cite{Mita25PRAP,Mita25arXiv} even at room temperatures. Although the current cavity magnonics still lacks the DSC regime, recent developments in achieving USC represent a significant step forward, offering novel platforms for the study of nonperturbative strong-coupling phenomena.

In this paper, we study nonperturbative strong-coupling, i.e., USC and DSC, in cavity magnonics systems in which coherent magnetization dynamics (magnons) in ferromagnets are coupled to microwave photons in a single-mode $LC$ resonator. 
There are two key concepts in this work. 
The first one is a nontrivial frequency shift in the eigenmode of MP, which emerges in the USC and DSC regimes. 
This frequency shift can be found even in a classical circuit model that accounts for the magnetization dynamics via the magnetic flux although the detail mechanism of the frequency shift remains unclear within the classical description. 
Based on the circuit model, we thereby derive a minimal quantum mechanical model for generic cavity magnonics, which is equivalent to a two-mode version of the Hopfield Hamiltonian and elucidates the mechanism behind the observed frequency shift. 
Also, we establish a connection between the frequency shift and several quantum quantities, such as the ground-state particle number, quantum fluctuations associated with Heisenberg's principle, and entanglement entropy. 
The second key concept is enhancement of these quantum quantities due to ``soft magnons'' whose excitation gap in the magnon dispersion vanishes at the critical magnetic field \cite{Yuan21,Bauer23}. Such soft magnons can be found in anisotropic (anti)ferromagnets with canted magnetization configurations induced by an external magnetic field applied perpendicular to the magnetic anisotropy axis \cite{Iihama14,de Wal23}.
By utilizing such an anisotropic ferromagnet, we  demonstrate that these quantum quantities are drastically enhanced and may even diverge at the critical field where the bare magnon frequency vanishes.

This paper is organized as follows: In Sec.~\ref{Classical model}, we introduce an effective circuit model that describes MPs for ferromagnets with a shape of rod, thin film, and sphere. Also, based on the circuit model, we discuss the eigenmode of MPs and emergence of a nontrivial frequency shift in the USC and DSC regimes. 
In Sec.~\ref{Quantum model}, we derive a minimal quantum mechanical model for generic cavity magnonics and analyze the mechanism of the frequency shift. 
Section~\ref{Quantum properties} focuses on quantum properties: we compute the ground-state particle number, two quadrature variances of the original boson mode, and entanglement entropy, and clarify their connection to the frequency shift.  We also demonstrate divergence of these quantum quantities in the framework of soft magnons.
Finally, in Sec.~\ref{Summary} we summarize the main findings and discuss the limitations and applicability of the proposed model.

\begin{figure*}[ptb]
\begin{centering}
\includegraphics[width=0.99\textwidth,angle=0]{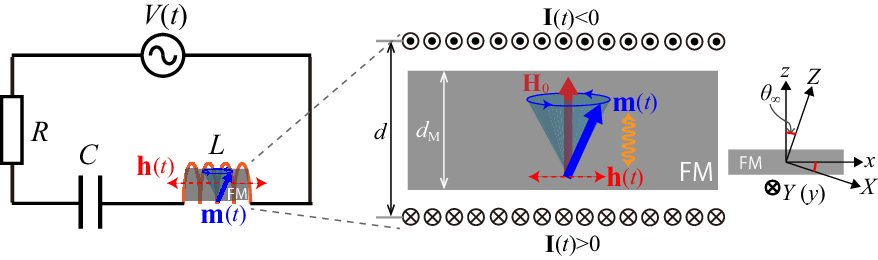} 
\par\end{centering}
\caption{
Effective circuit model of cavity magnonics systems in which ${\bf h}(t)$ is a microwave magnetic field (photon) in an inductor and ${\bf m}(t)$ is uniform magnetization dynamics (magnon) in a ferromagnet (FM) biased by an external static magnetic field ${\bf H}_0$. Here, the cross-section of the inductor is assumed to be circuital, which is characterized by $d$ and $d_{\rm M}$ being diameters of the inductor and ferromagnet, respectively. Depending on the balance between the external magnetic field and demagnetization field, the equilibrium position of the magnetization is determined, which is characterized by 
a polar angle $\theta_{\infty}$ measured from the $z$-axis.}
\label{Fig:system}
\end{figure*}

\section{Classical description of magnon-polariton}\label{Classical model}

\subsection{Effective circuit model}

First, we describe cavity magnonics systems in the thermodynamic limit $N_{\rm S} \gg 1$ ($N_{\rm S}$ is the total spin number of ferromagnets) by using an effective circuit model \cite{Bai15,Grigoryan18,Chiba24APL,Chiba24MSJ},
which is schematically shown in Fig.~\ref{Fig:system}. 
An effective $RLC$-circuit describes a microwave photon system of a single $LC$ resonator and a uniform magnetization dynamics of a ferromagnet inserted in an inductor describes a coherent magnon system. Inside the inductor, the magnetization of the inserted ferromagnet is excited by a microwave magnetic field, denoted by ${\bf H}(t)$, with a resonant mode of the $RLC$-circuit. The circuit mode is coupled to the Kittel mode of magnetization dynamics, ${\bf M}(t)$, via the magnetic-dipole (Zeeman) interaction, which is responsible for the coherent magnon-photon coupling. 

The magnetization dynamics of the ferromagnet is described by the Landau-Lifshitz-Gilbert (LLG) equation
\begin{align}
\frac{d \bf m}{d t} = -\gamma{\bf m}\times\left(  -\frac{1}{M_{\rm s}}\frac{\delta U_{\rm m}}{\delta{\bf m}} +\mu_0k_{\rm c}{\bf H}(t)\right) + \alpha{\bf m}\times\frac{d \bf m}{d  t}
,\label{LLG}
\end{align}
where ${\bf m}(t) = {\bf M}(t)/M_{\rm s}$ is the unit vector along the magnetization direction of a ferromagnet with the saturation magnetization $M_{\rm s}$, $\gamma$ is the gyromagnetic ratio, $\mu_0$ is the permeability of free space, $k_{\rm c}$ is the Nagaoka coefficient that is responsible for the mode volume of the microwave photon inside the inductor, and $\alpha$ is the intrinsic Gilbert damping constant. 
Here, the total magnetic energy is assumed by
\begin{align}
U_{\rm m} = - \mu_0M_{\rm s}{\bf m}\cdot{\bf H}_0 - \frac{1}{2}\mu_0M_{\rm s}{\bf m}\cdot{\bf H}_{\rm d}[{\bf m}]
,\label{Um}
\end{align}
where ${\bf H}_0 = H_0\hat{\bf z}$ is the external static magnetic field and ${\bf H}_{\rm d}[{\bf m}]$ describes the demagnetization field due to the shape magnetic anisotropy depending on shapes of a ferromagnet such as rod, thin film, and sphere. In Eq.~\eqref{Um}, the exchange interaction term is neglected. This approximation is valid when the system scales of the ferromagnet are much larger than the exchange length \cite{Hoffman13,Cornelissen16}. For instance, yttrium iron garnet (YIG), used for calculations in this paper, possesses a relatively long exchange length on the order of 10~nm under low external magnetic fields \cite{Cornelissen16}. In fact, typical cavity magnonics system operates in the microwave regime whose wavelength is much larger than the exchange length.

Inside the inductor, the microwave magnetic field is induced by an alternating current $I(t)$ which is governed by the $RLC$-circuit equation (Kirchhoff's voltage law)
\begin{align}
N\frac{d \Phi(t)}{dt} + RI(t) + \frac{Q(t)}{C} = V(t)
,\label{KVL}
\end{align}
where $N$ is a turn number, $\Phi(t)$ is the magnetic flux inside the inductor, $Q(t)$ is the accumulated charge on the capacitor with an electrostatic capacitance $C$, $R$ is the resistance, and $V(t) = V_0e^{-i\omega t}$ is an applied ac voltage with an angular frequency $\omega$ (experimentally due to input signal from the vector network analyzer). According to Amp\`{e}re's law, the microwave magnetic field is given by ${\bf H}(t) = (NI(t)/l)\hat{\bf x}$, where $l$ is the inductor length. Then, the magnetic flux is 
\begin{align}
\Phi(t) = \mu_0M_{\rm s}\left(  k_{\rm c}{\bf h}(t) + k_{\rm m}\eta_{\rm S}{\bf m}(t)\right)\cdot{\bf S}
,\label{magnetic flux}
\end{align}
where ${\bf h}(t) = {\bf H}(t)/M_{\rm s}$ along the $x$-axis, ${\bf S} = S\hat{\bf x}$ with $S$ being the cross-section area of the inductor, $k_{\rm m}$ is the Nagaoka coefficient of a ferromagnet, and $\eta_{\rm S}$~($0\leq\eta_{\rm S}\leq1$) represents the volume ratio between the inductor and ferromagnet. 
Defining the circuit mode angular frequency by $\omega_{\rm c} = 1/\sqrt{LC}$ with the inductance $L = k_{\rm c}\mu_0N^2S/l$, Eq.~(\ref{KVL}) is equivalent to the equation of motion for a single harmonic oscillator
\begin{align}
\left(  \frac{d^2}{dt^2} + 2\beta\omega_{\rm c}\frac{d}{dt} + \omega_{\rm c}^2\right)h_x + k_{\rm m}\eta_{\rm S}\frac{d^2m_x}{dt^2}
= \omega_{\rm c}^2h_V
,\label{Maxwell}
\end{align}
where $\beta = R\sqrt{C/L}/2$ is an effective damping of the circuit and $h_V = NC/(lM_{\rm s})dV/dt$.

\subsection{Shape-dependence of ferromagnet}

Since the magnon eigenmodes are significantly influenced by the shape-dependent magnetic anisotropy [Eq.~(\ref{Um})], we begin by evaluating the magnon eigenfrequencies for representative ferromagnetic geometries, including rod, thin film, and spherical shapes. This arrangement might enhance the applicability of the proposed model to cavity magnonics systems with various shape of magnet.

\subsubsection{Cylindrical rod shape}

We consider a ferromagnet with a cylindrical rod shape which generates a shape-based uniaxial magnetic anisotropy, as shown below. Then, the demagnetization field in Eq.~(\ref{Um}) is expressed by 
\begin{align}
{\bf H}_{\rm d}[{\bf m}] = - \frac{1}{2}M_{\rm s}\left(  m_y\hat{\bf y} + m_z\hat{\bf z}\right)
.\label{Hdrod}
\end{align}
Introducing the polar angle $\theta(t)$ and azimuthal angle $\varphi(t)$ for ${\bf m}(t) = (\cos\varphi\sin\theta, \sin\varphi\sin\theta, \cos\theta)$, Eq.~(\ref{Um}) can be expressed as
$U_{\rm m}/(\mu_0M_{\rm s}^2) = - (H_0/M_{\rm s})\cos\theta + \left( 1 - \cos^2\varphi\sin^2\theta\right)/4$,
giving the equilibrium position of ${\bf m}(t \to \infty)$ in the absence of ${\bf H}(t)$ by
\begin{align}
\left(\theta_{\infty}, \varphi_{\infty}\right) = \left(\cos^{-1}\frac{2H_0}{M_{\rm s}}, 0\right)
.\label{thetainfrod}
\end{align}
In the range of $H_0 < M_{\rm s}/2$, the magnetic potential has two-degenerate equilibrium positions along the $\varphi_{\infty}=0$ direction, leading to canted magnetization configurations (spontaneous $Z_2$-symmetry breaking) induced by the external magnetic field. For $H_0 \geq M_{\rm s}/2$, a strong external field aligns the magnetization to the $z$ direction, thereby setting $\theta_{\infty} = 0$.

It is convenient to consider the magnetization dynamics in the $XYZ$-coordinate system [see Fig.~\ref{Fig:system}], in which the magnetization is stabilized along the $Z$-axis determined by Eq.~(\ref{thetainfrod}). Denoting the transformation matrix as $\mathcal{R}(\theta_{\infty})$ around the $Y(y)$-axis, in the presence of ${\bf H}(t)$, the magnetization ${\bf n}(t) = \mathcal{R}(\theta_{\infty}){\bf m}(t) = (n_X,n_Y,n_Z)$ precesses around ${\bf H}_{\mathcal{R}} = \mathcal{R}(\theta_{\infty})\left(  -\delta U_{\rm m}/\delta{\bf m}\right)/(\mu_0M_{\rm s})  = (H_X,0,H_Z)$, where
$H_{X} = -H_0\sin\theta_{\infty}
+ M_{\rm s}\left(  n_X\cos\theta_{\infty} + n_Z\sin\theta_{\infty}\right)\cos\theta_{\infty}/2$ and 
$H_{Z} = H_0\cos\theta_{\infty} 
+ M_{\rm s}\left(  n_X\cos\theta_{\infty} + n_Z\sin\theta_{\infty}\right)\sin\theta_{\infty}/2$.
Then, the LLG equation in the $XYZ$-coordinate system is described by
\begin{align}
\frac{d \bf n}{d t} = -\gamma{\bf n}\times\left(  \mu_0{\bf H}_{\mathcal{R}} + \mu_0M_{\rm s}k_{\rm c}{\bf p}(t)\right) + \alpha{\bf n}\times\frac{d \bf n}{d t}
,\label{LLGn}
\end{align}
where ${\bf p}(t) = \mathcal{R}(\theta_{\infty}){\bf h}(t) = (p_X(t),0,p_Z(t))$. 

For a small input from the microwave source $V(t)$, the magnetization dynamics can be linearized as ${\bf n}(t) = (n_X(t),n_Y(t),1) \propto {\bf p}(t)$ with $|n_X|,|n_Y| \ll 1$. 
Then, the linearized LLG equation in the $XYZ$-system becomes the equation of motion for a harmonic oscillator
\begin{align}
\left(  \frac{d^2 }{d t^2}
+ 2\alpha_{\rm m}\omega_{\rm m}\frac{d }{d t} 
+ \omega_{\rm m}^2\right)n_X 
- k_{\rm c}\omega_{\rm M}\omega_{\parallel1}p_X 
= \omega_{\parallel1}\omega_{\perp}
,\label{LLGnXloscillator}
\end{align}
where 
\begin{align}
\omega_{\rm m} 
&= \sqrt{\omega_{\parallel1}\omega_{\parallel2}}
,\label{omegam}\\
\alpha_{\rm m}
&= \frac{\alpha}{\sqrt{1 + \alpha^2}}\frac{\omega_{\parallel1} + \omega_{\parallel2}}{2\sqrt{\omega_{\parallel1}\omega_{\parallel2}}}
,\label{alphaeff}
\end{align}
are the magnon eigenfrequency and the effective damping constant of the coherent magnon, respectively, and $\omega_{\rm M} = \gamma'\mu_0M_{\rm s}$ with $\gamma' = \gamma/\sqrt{1 + \alpha^2}$. Here, we define 
$\omega_{\parallel1}
= \gamma'\mu_0\left(  H_0\cos\theta_{\infty} 
+ M_{\rm s}\sin^2\theta_{\infty}/2\right)$, 
$\omega_{\parallel2}
= \gamma'\mu_0\left(  H_0\cos\theta_{\infty} 
- M_{\rm s}\cos2\theta_{\infty}/2\right)$,
and
$\omega_{\perp}
= \gamma'\mu_0\left(  -H_0\sin\theta_{\infty}
+ M_{\rm s}\sin2\theta_{\infty}/4\right)$.
Therefore, the magnon eigenfrequency of the rod shape is 
\begin{align}
\omega_{\rm m} = 
\begin{cases}
 \gamma'\mu_0\sqrt{\left(  M_{\rm s}/2\right)^2 - H_0^2} & (H_0 < M_{\rm s}/2) \\
 \gamma'\mu_0\sqrt{H_0\left(  H_0 - M_{\rm s}/2\right)} & (H_0 \geq M_{\rm s}/2)
\end{cases}
.\label{omegamrod}
\end{align}
At $H_0 = M_{\rm s}/2$, the magnon eigenfrequency once becomes zero, indicating the behavior of  soft magnons \cite{Yuan21,Bauer23}. 

On the other hand, Eq.~(\ref{Maxwell}) in the $XYZ$-system becomes 
\begin{align}
&\left(  \frac{d^2}{dt^2} + 2\beta\omega_{\rm c}\frac{d}{dt} + \omega_{\rm c}^2\right)p_X + k_{\rm m}\eta_{\rm S}\frac{d^2n_X}{dt^2}\cos^2\theta_{\infty} = \omega_{\rm c}^2p_V^X
,\label{Maxwelln}
\end{align}
where ${\bf p}_V(t) = \mathcal{R}(\theta_{\infty}){\bf h}_V(t)$
in the framework of the linearized magnetization dynamics.

\subsubsection{Thin film shape}

Assuming a thin film of ferromagnets, the demagnetization field in Eq.~(\ref{Um}) is expressed by 
\begin{align}
{\bf H}_{\rm d}[{\bf m}] = - M_{\rm s}m_z\hat{\bf z}
.\label{Hdfilm}
\end{align}
Then, in the polar-coordinate Eq.~(\ref{Um}) is described as
$U_{\rm m}/(\mu_0M_{\rm s}^2) = - (H_0/M_{\rm s})\cos\theta + \cos^2\theta/2$,
which gives the equilibrium position of ${\bf m}(t \to \infty)$ in the absence of ${\bf H}(t)$ by
\begin{align}
\left(\theta_{\infty}, \varphi_{\infty}\right) = \left(\cos^{-1}\frac{H_0}{M_{\rm s}}, \varphi_0\right)
,\label{thetainffilm}
\end{align}
where $\varphi_0 \in [0,2\pi]$. This arbitrariness of $\varphi_0$ indicates that the magnetic potential $U_{\rm m}$ has the $U(1)$ symmetry, which generates a zero-mode of the magnon excitation due to the spontaneous symmetry breaking in the low-field regime ($H_0 < M_{\rm s}$). For $H_0 \geq M_{\rm s}/2$, the magnetization is aligned to the $z$ direction, resulting in $\theta_{\infty} = 0$.

For simplicity (but without loss of generality), we choose the phase of $\varphi_0 = 0$. As in Fig.~\ref{Fig:system}~(b), it is convenient to consider the magnetization dynamics in the $XYZ$-coordinate system, in which magnetization is stabilized along the $Z$-axis determined by Eq.~(\ref{thetainffilm}). Then, a coherent magnon system is described by Eq.~(\ref{LLGnXloscillator}) with
$\omega_{\parallel1}
= \gamma'\mu_0\left(  H_0\cos\theta_{\infty} 
- M_{\rm s}\cos^2\theta_{\infty}\right)$, 
$\omega_{\parallel2}
= \gamma'\mu_0\left(  H_0\cos\theta_{\infty} 
- M_{\rm s}\cos2\theta_{\infty}\right)$, 
$\omega_{\perp}
= \gamma'\mu_0\left(  -H_0\sin\theta_{\infty}
+ M_{\rm s}\sin2\theta_{\infty}/2\right)$.
Therefore, the magnon eigenfrequency of the thin film shape is 
\begin{align}
\omega_{\rm m} = 
\begin{cases}
 ~0 & (H_0 < M_{\rm s}) \\
 \gamma'\mu_0\left(  H_0 - M_{\rm s}\right) & (H_0 \geq M_{\rm s})
\end{cases}
.\label{omegamfilm}
\end{align}

On the other hand, in the $XYZ$-coordinate a microwave photon system is described by Eq.~(\ref{Maxwelln}) in the framework of the linearized magnetization dynamics.

\subsubsection{Spherical shape}

Assuming a spherical shape of ferromagnets, the demagnetization field in Eq.~(\ref{Um}) is expressed by 
\begin{align}
{\bf H}_{\rm d}[{\bf m}] = - \frac{1}{3}M_{\rm s}{\bf m}
.\label{Hdsphere}
\end{align}
Then, in the polar-coordinate Eq.~(\ref{Um}) is described as
$U_{\rm m}/(\mu_0M_{\rm s}^2) = - (H_0/M_{\rm s})\cos\theta + 1/6$,
which gives the equilibrium position of ${\bf m}(t \to \infty)$ in the absence of ${\bf H}(t)$ by
\begin{align}
\left(\theta_{\infty}, \varphi_{\infty}\right) = \left(0, \varphi_0\right)
,\label{thetainffilm}
\end{align}
where $\varphi_0 \in [0,2\pi]$. 

For simplicity, we choose the phase of $\varphi_0 = 0$. Then, a coherent magnon system is described by Eq.~(\ref{LLGnXloscillator}) with
$\omega_{\parallel1} = \omega_{\parallel2} = \gamma'\mu_0H_0$ and $\omega_{\perp} = 0$.
Therefore, the magnon eigenfrequency of the spherical shape is given by
\begin{align}
\omega_{\rm m}
&= \gamma'\mu_0H_0
.\label{omegamsphere}
\end{align}

Since the magnetization dynamics in spherical ferromagnets is isotropic, a microwave photon system is simply described by Eq.~(\ref{Maxwell}).

\subsection{Eigenmodes of magnon-polariton}

Here, we investigate eigenmodes of MP (polariton eigenfrequency) in the framework of the classical circuit model.
Introducing effective charge and voltage, $p_X = \omega_{\rm M}^{-1}d q_X/dt$ and $p_V^X = \omega_{\rm M}^{-1}d v_X/dt$, Eqs.~(\ref{LLGnXloscillator}) and (\ref{Maxwelln}) are transformed to
\begin{align}
&m_1\left(  \frac{d^2}{dt^2} + 2\beta\omega_{\rm c}\frac{d}{dt} + \omega_{\rm c}^2\right)q_X + \lambda^2\frac{d N_X}{dt} = m_1\omega_{\rm c}^2v_X(t)
,\label{MaxwellqX}\\
&m_2\left(  \frac{d^2 }{d t^2}
+ 2\alpha_{\rm m}\omega_{\rm m}\frac{d }{d t} 
+ \omega_{\rm m}^2\right)N_X
- \lambda^2\frac{d q_X}{dt} = 0
,\label{LLGnX}
\end{align}
where $m_{1} = k_{\rm c}\omega_{\rm \parallel1}$ and $m_{2} = k_{\rm m}\eta_{\rm S}\omega_{\rm M}\cos^2\theta_{\infty}$ are effective masses (in units of angular frequency), $N_X = n_X - \omega_{\parallel1}\omega_{\perp}/\omega_{\rm m}^2$ is the effective magnetic flux (magnetization), and 
\begin{align}
\lambda = \sqrt{m_1m_2} = \sqrt{k_{\rm c}k_{\rm m}\eta_{\rm S}\omega_{\rm \parallel1}\omega_{\rm M}}\cos\theta_{\infty}
.\label{lambda}
\end{align}
Assuming harmonic solutions $q_X(t) = \tilde{q}_Xe^{-i\omega t}$ and $N_X(t) = \tilde{N}_Xe^{-i\omega t}$ for the input $v_X(t) = v_0e^{-i\omega t}$, we obtain the coupled LLG and $RLC$-circuit equations in the frequency domain
\begin{align}
\bar{\Omega}
\begin{pmatrix}
\tilde{q}_X \\
\tilde{N}_X
\end{pmatrix}
=
\begin{pmatrix}
-m_1\omega_{\rm c}^2v_0 \\
0
\end{pmatrix}
\label{LLG-RLC}
\end{align}
with 
\begin{align}
\bar{\Omega} = 
\begin{pmatrix}
m_1\left(  \omega^2 + 2i\beta\omega_{\rm c}\omega - \omega_{\rm c}^2\right)
& i\lambda^2\omega \\
-i\lambda^2\omega
& m_2\left(  \omega^2
+ 2i\alpha_{\rm m}\omega_{\rm m}\omega  
- \omega_{\rm m}^2\right)
\end{pmatrix}
.\label{OmegaMP}
\end{align}
To investigate an influence of damping on the hybridized MP modes, we first calculate the transmission amplitude using input-output formalism from Eq.~(\ref{LLG-RLC}), 
\begin{align}
S_{21} = \Gamma\frac{q_X}{v_0}
 = \Gamma\frac{-m_1m_2\omega_{\rm c}^2\left(  \omega^2
+ 2i\alpha_{\rm m}\omega_{\rm m}\omega  
- \omega_{\rm m}^2\right)}{{\rm det}~\bar{\Omega}}
,\label{S21}
\end{align}
where $\Gamma = 2\beta$ determines the cavity$|$cable impedance mismatch \cite{Bai15,Grigoryan18}.

To obtain the eigenfrequency of the hybridized MP modes, we neglect the damping parameters in Eq.~(\ref{OmegaMP}) by setting $\alpha_{\rm m} = \beta = 0$.
Hence, hybridized eigenmodes are calculated by solving the determinant of Eq.~(\ref{OmegaMP}), leading to [see also Appendix~\ref{Hamilton eq}]
\begin{align}
\omega_\pm^2
& = \frac{\omega_{\rm c}^2 + \omega_{\rm m}^2 + \lambda^2}{2} \pm \sqrt{\frac{\left(  \omega_{\rm c}^2 + \omega_{\rm m}^2 + \lambda^2\right)^2}{4} - \omega_{\rm c}^2\omega_{\rm m}^2}
.\label{omegaNRWA}
\end{align}
Then, the coupling strength at the original modes crossing point ($\omega_{\rm m} = \omega_{\rm c}$) is given by
\begin{align}
g \equiv \frac{\omega_+ - \omega_-}{2}\bigg|_{\omega_{\rm m} = \omega_{\rm c}}
= \frac{\lambda}{2}
.\label{gNRWA}
\end{align}
For further discussion, we introduce a classical version of the rotating-wave approximation (RWA) \cite{Chiba24APL,Chiba24MSJ}, that is, $\omega^2 - \omega_{\rm c}^2 \approx 2\omega_{\rm c}\left(  \omega - \omega_{\rm c}\right)$ and $\omega^2 - \omega_{\rm m}^2 \approx \left(  \omega_{\rm c} + \omega_{\rm m}\right)\left(  \omega - \omega_{\rm m}\right)$,  to the determinant of Eq.~(\ref{OmegaMP}). Also, writing $\omega = \omega_{\rm c} + \delta$ with $\delta \ll \omega$, we neglect the terms of order $\delta^2$ and $\delta \lambda$, leading to
\begin{align}
\omega_\pm^{\rm RWA} = \frac{1}{2}\left[  \omega_{\rm c} + \omega_{\rm m} \pm \sqrt{\left(  \omega_{\rm c} - \omega_{\rm m}\right)^2 + \frac{2\lambda^2\omega_{\rm c}}{\omega_{\rm c} + \omega_{\rm m}}}\right]
.\label{omegaRWA}
\end{align}
Therefore, the coupling strength at the original modes crossing point is given by
\begin{align}
g^{\rm RWA} \equiv \frac{\omega_+^{\rm RWA} - \omega_-^{\rm RWA}}{2}\Bigg|_{\omega_{\rm m} = \omega_{\rm c}}
= g
,\label{gRWA}
\end{align}
which indicates that the RWA does not change the magnitude of the coupling strength (the Rabi-like splitting).

Hereafter, we focus on a cylindrical rod shape of  ferromagnets by assuming that the cross-sections of the inductor and the ferromagnet are circular, which is characterized by $d$ and $d_{\rm M}$ being diameters of the inductor and ferromagnet, respectively. Then, we have the relation $\eta_{\rm S} = d_{\rm M}^2/d^2$. Throughout this paper, we use circuit parameters as the following: $R = 1$~$\Omega$, $C = 1$~pF, and $L = 6.2$~nH (for $N = 5$, $l = 15$~mm, and $d = 2$~mm), which corresponds to $k_{\rm c} = 0.947$, $\omega_{\rm c}/(2\pi) = 2.0$~GHz, and $\beta = 6.3\times10^{-3}$. Also, for simplicity, the Nagaoka coefficient of the ferromagnet is assumed to be equivalent to that of the inductor, $k_{\rm m} = k_{\rm c}$.
The material parameters for the ferromagnet are $\gamma = 1.76\times10^{11}$~${\rm T^{-1}s^{-1}}$, $\alpha = 10^{-4}$, and $M_{\rm s} = 1.6\times10^5$~Am$^{-1}$~($\mu_0M_{\rm s} \approx 200$~mT) by assuming ${\rm Y_3Fe_5O_{12}}$ \cite{Mita25PRAP,Bai15}. 

\begin{figure}[hptb]
\begin{centering}
\includegraphics[width=0.45\textwidth,angle=0]{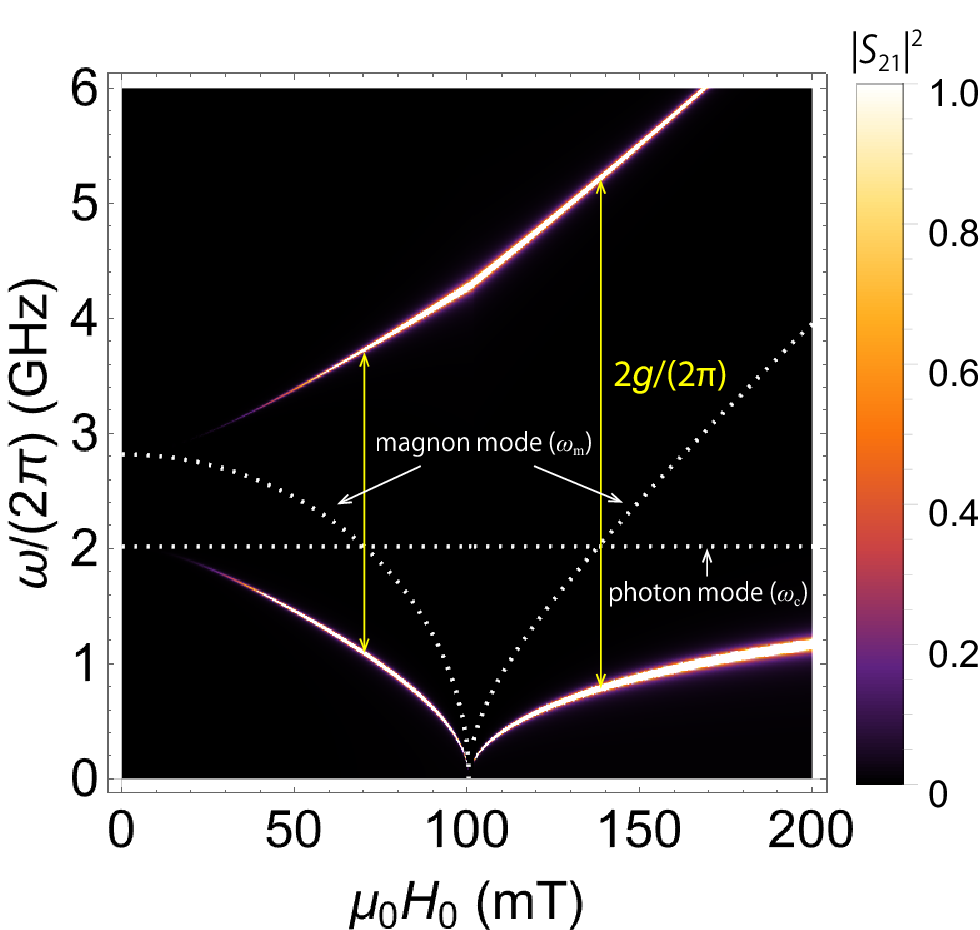}
\par\end{centering}
\caption{
Calculated transmission amplitude ($|S_{21}|^2$) of the MP hybridized modes as functions of an input frequency ($\omega/(2\pi)$) and an external magnetic field ($\mu_0H_0$) for $d_{\rm M}/d = 1$. Asymmetric Rabi-like splittings emerges at the original mode crossing points, which is an experimental signature of the nonperturbative strong coupling.
}
\label{Fig:S21}
\end{figure}

\begin{figure}[hptb]
\begin{centering}
\includegraphics[width=0.48\textwidth,angle=0]{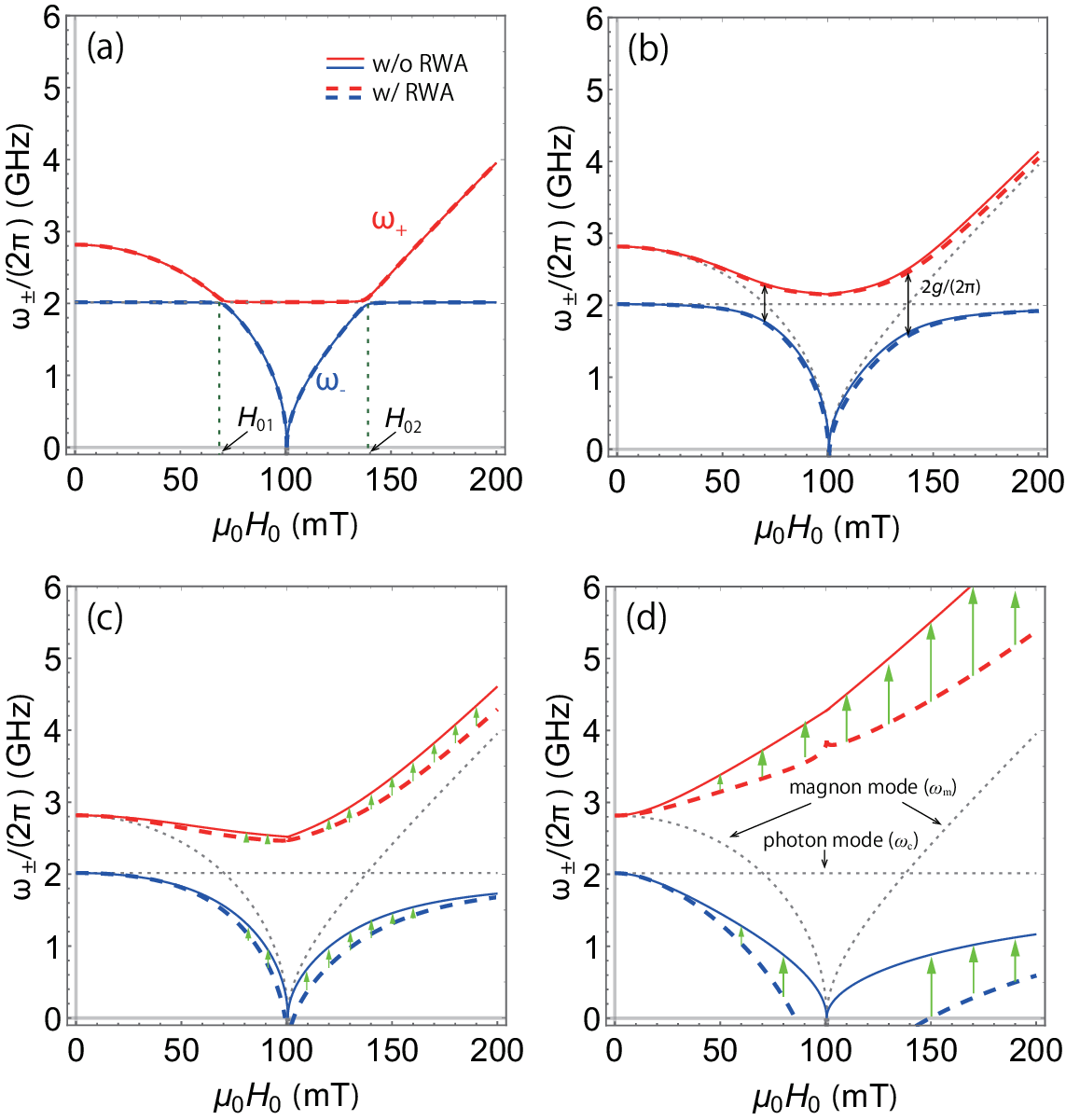}
\par\end{centering}
\caption{
Calculated eigenfrequency ($\omega_\pm$) of the MP hybridized modes as a function of an external magnetic field ($\mu_0H_0$) for different values of $d_{\rm M}/d$: (a) 0.02, (b) 0.2, (c) 0.4, and (d) 1. The solid lines are results of the no RWA case that is equivalent to the eigenfrequency of the two-mode Hopfield model ($\omega_\pm^{\rm Hopfield}$ in Eq.~(\ref{omegaHopfield})). Green arrows represent positive frequency shifts at each external magnetic field.
}
\label{Fig:omegapm}
\end{figure}

First, we consider the influence of the system damping ($\alpha_{\rm m}$ and $\beta$) on hybridized MP modes. In Fig.~\ref{Fig:S21}, we plot the transmission amplitude $|S_{21}|^2$ of the hybridized MP modes for $d_{\rm M}/d = 1$ with $\alpha = 10^{-4}$ and $\beta = 6.3\times10^{-3}$. This is just an example of MP in the nonperturbative strong-coupling regime that is a main subject of this paper. As seen, even in the presence of the system damping one can clearly observe the hybridized MP modes on the transmission amplitude, which corresponds to the eigenfrequency displayed in Fig.~\ref{Fig:omegapm}~(d) (the detail explanation will be given in the next paragraph). Since the hybridized MP modes are well defined in the sense of an experimentally observable quantity, hereafter we concentrate on the eigenfrequency of the hybridized MP modes by disregarding the system damping. Also, in Fig.~\ref{Fig:S21}, one can find asymmetric Rabi-like splitting at the original mode crossing points, implying that the hybridized system goes into the nonperturbative strong-coupling regime characterized by a nontrivial frequency shift as discussed below.

\begin{figure}[hptb]
\begin{centering}
\includegraphics[width=0.45\textwidth,angle=0]{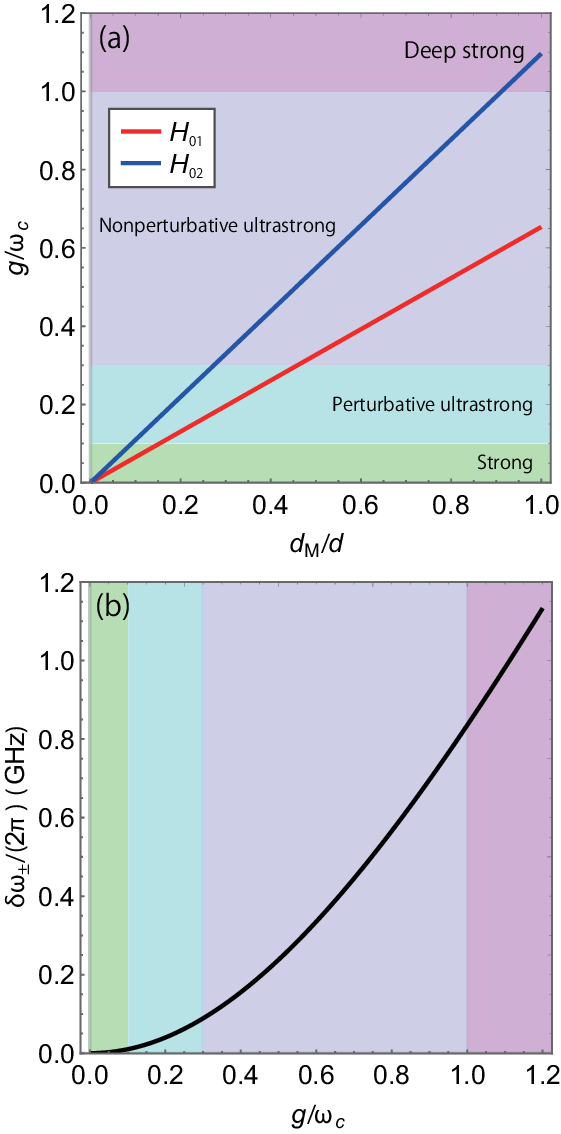}
\par\end{centering}
\caption{
(a) Calculated coupling ratios ($g/\omega_{\rm c}$) of the hybridized MP modes as a function of $d_{\rm M}/d$ for different values of an external magnetic field. For this calculation, $d = 2$~mm is fixed.
(b) Calculated frequency shift ($\delta\omega_\pm$) at the original modes crossing point as a function of $g/\omega_{\rm c}$.
}
\label{Fig:g/wc}
\end{figure}

Figure~\ref{Fig:omegapm}~(a) shows the calculated eigenfrequency of the hybridized MP modes for $d_{\rm M}/d = 0.02$, from which the estimated coupling ratios $g/\omega_{\rm c}$ are 0.01 at $H_{01}$ and 0.02 at $H_{02}$. Here, $H_{01} = 70$~mT and $H_{02} = 138$~mT are values of the external magnetic field at the original mode crossing points [see in Fig.~\ref{Fig:omegapm}~(a)]. At the two points, the condition of $g/\omega_{\rm c} > \alpha_{\rm m},\beta$ but $g/\omega_{\rm c} < 0.1$ indicate that the MPs for $d_{\rm M}/d = 0.02$ reside in the strong coupling regime. Also, in Fig.~\ref{Fig:omegapm}~(a) one can see a good agreement between the RWA and no RWA cases. 
Figure~\ref{Fig:omegapm}~(b) shows the calculated eigenfrequency of the MPs for $d_{\rm M}/d = 0.2$, which gives $g/\omega_{\rm c} = 0.13$ at $H_{01}$ and $g/\omega_{\rm c} = 0.22$ at $H_{02}$. So, these coupling ratios  satisfied with $g/\omega_{\rm c} > 0.1$ indicate that the MPs go into the USC regime wheres the result of no RWA slightly deviates from that of RWA. On the context, this regime is usually referred to be ``perturbative'' USC in the cavity QED.
In contrast, as shown in Fig.~\ref{Fig:omegapm}~(c), the result of no RWA for $d_{\rm M}/d = 0.4$ is clearly different from that of RWA, whose difference is represented by green arrows accompanied with a positive frequency shift. Also, the estimated coupling ratios are $g/\omega_{\rm c} = 0.26$ at $H_{01}$ and $g/\omega_{\rm c} = 0.44$ at $H_{02}$. On the context, this regime corresponds to the ``nonperturbative'' USC regime in the cavity QED \cite{Diaz19,Kockum19}. 
For $d_{\rm M}/d = 1$, i.e., in the fully filled case, the calculated hybridized MP modes are shown in Fig.~\ref{Fig:omegapm}~(d), where the coupling ratios are $g/\omega_{\rm c} = 0.65$ at $H_{01}$ and $g/\omega_{\rm c} = 1.1$ at $H_{02}$. Especially, the later one satisfied with $g/\omega_{\rm c} > 1$ reaches the DSC regime, in which the hybridized MP modes is far away from the original photon and magnon modes. Also, in this regime, the lower MP modes obtained by the RWA become negative values around 100~mT (not fully shown), which indicates that the RWA clearly breaks down and therefore exhibits nontrivial properties. In fact, as found in Fig.~\ref{Fig:omegapm}~(c) and more brightly in Fig.~\ref{Fig:omegapm}~(d), larger positive frequency shifts occur over the wide range of the external magnetic field. These positive frequency shifts are nontrivial. This is because the (vacuum) BS shift induced by the USC with no RWA is expected to be a negative frequency shift \cite{Diaz10,Li18} due to the coherent coupling, which destabilizes the ground state of polaritons and leads to the superradiant phase transition.

Figure~\ref{Fig:g/wc}~(a) shows the calculated coupling ratios of the hybridized MP modes as a function of $d_{\rm M}/d$ for $H_{01} = 70$~mT and $H_{02} = 138$~mT. For this calculation, we adopt the same parameters for the circuit and the ferromagnet used in Fig.~\ref{Fig:omegapm}. Since the shape of the ferromagnet is a cylindrical rod, we have the relation $\eta_{\rm S} = d_{\rm M}^2/d^2$. Then, Eq.~(\ref{gNRWA}) leads to the relation $g \propto d_{\rm M}/d$, which can be seen in Fig.~\ref{Fig:g/wc}, i.e., the coupling ratios at both $H_{01}$ and $H_{02}$ linearly increase with increasing $d_{\rm M}/d$. Remarkably, for the both cases, the nonperturbative USC is achieved in the wide parameter range of $d_{\rm M}/d$. In contrast, the DSC regime is generated only by the $H_{02}$ point for $d_{\rm M}/d > 0.91$. This difference reflects the $\mu_0H_0$-dependence of the coupling strength via Eq.~(\ref{lambda}).
Figure~\ref{Fig:g/wc}~(b) shows the calculated frequency shift at the original modes crossing point, which is defined by 
\begin{align}
\delta\omega_\pm
&\equiv \omega_\pm - \omega_\pm^{\rm RWA}
= \omega_{\rm c}\left[  \sqrt{1 + \left(  \frac{g}{\omega_{\rm c}}\right)^2} - 1\right] \geq 0
.\label{omega shift}
\end{align}
This frequency shift corresponds to the positive frequency shifts at the original modes crossing point in Fig.~\ref{Fig:omegapm}~(c) and Fig.~\ref{Fig:omegapm}~(d). As seen in Fig.~\ref{Fig:g/wc}~(b), the calculated frequency shift starts to gradually increase when the MP system goes into the perturbative USC regime, whose tendency is similar to that of the BS shift \cite{Diaz19}. However, the sign of the frequency shift is opposite to that of the BS shift, as discussed in Fig.~\ref{Fig:omegapm}, and hence the detail mechanism of the frequency shift still remains unclear within the classical description.

\section{Quantum-classical correspondence}
\label{Quantum model}

\subsection{Minimal quantum mechanical model}

To elucidate the mechanism of the positive frequency shift found in the classical circuit model, our first step is to derive a minimal quantum mechanical model for the cavity magnonics system described in Sec~\ref{Classical model}. By setting $\alpha_{\rm m} = \beta = 0$, the equations of motion~(\ref{MaxwellqX}) and (\ref{LLGnX}) can be derived from the Lagrange equation with the Lagrangian of the MP,  $\mathcal{L} = \mathcal{L}_0 + \mathcal{L}_{\rm ex}$, where
\begin{align}
\mathcal{L}_0
=& \frac{m_1}{2}\dot{q}_X^2 - \frac{m_1\omega_{\rm c}^2}{2}\left( 
 q_X - v_X(t)\right)^2\nonumber\\
&+ \frac{m_2}{2}\dot{N}_X^2 - \frac{m_2\omega_{\rm m}^2}{2}N_X^2
 ,\label{L0}\\
\mathcal{L}_{\rm ex}
=&  \lambda^2\dot{q}_XN_X
.\label{Lex}
\end{align}
Note that in $\mathcal{L}_0$ the term $\propto \left(v_X(t)\right)^2$ corresponds to the energy of the microwave source $V(t)$. Introducing the canonical momenta: 
\begin{align}
\phi_X = \frac{\partial \mathcal{L}}{\partial \dot{q}_X} 
= m_1\dot{q}_X + \lambda^2 N_X
,~
P_X = \frac{\partial \mathcal{L}}{\partial \dot{N}_X}
= m_2\dot{N}_X
,\label{canonical momenta}
\end{align}
the usual quantization procedure  [see  Appendix~\ref{Hamilton eq}] leads to the Hamilton operator, $\hat{H} = \hat{H}_0 + \hat{H}_{\rm ex} + \hat{H}_{\rm dr}$, where
\begin{align}
\hat{H}_0
&= \frac{\hat{\phi}_X^2}{2m_1} + \frac{m_1\omega_{\rm c}^2}{2}\hat{q}_X^2+ \frac{\hat{P}_X^2}{2m_2}
 + \frac{m_2\omega_{\rm m}^2}{2}\hat{N}_X^2
,\label{QH0}\\
\hat{H}_{\rm ex}
&= -\lambda^2\frac{\hat{\phi}_X\hat{N}_X}{m_1} + \lambda^4\frac{\hat{N}_X^2}{2m_1}
,\label{QHex}\\
\hat{H}_{\rm dr}
&= - m_1\omega_{\rm c}^2\hat{q}_Xv_X(t)
.\label{QHdr}
\end{align}
Here, the commutation relations $[  \hat{q}_X,\hat{\phi}_X] = [  \hat{N}_X,\hat{P}_X] = i\hbar$ with the Dirac constant $\hbar$ are introduced. Note that we neglect the term $\propto \left(v_X(t)\right)^2$ the energy of the microwave source. 
Next, we move to the Fock-representation based on the following photon (magnon) creation $\hat{a}^{\dagger}~(\hat{b}^{\dagger})$ and annihilation $\hat{a}~(\hat{b})$ operators that obey
\begin{align}
\begin{pmatrix}
\hat{a} \\
\hat{a}^{\dagger}
\end{pmatrix}
= 
\frac{1}{\sqrt{2\hbar m_1\omega_{\rm c}}}
\begin{pmatrix}
m_1\omega_{\rm c} & i \\
m_1\omega_{\rm c} & -i 
\end{pmatrix}
\begin{pmatrix}
\hat{q}_X \\
\hat{\phi}_X
\end{pmatrix},\label{ahat}
\end{align}%
\begin{align}
\begin{pmatrix}
\hat{b} \\
\hat{b}^{\dagger}
\end{pmatrix}
= 
\frac{1}{\sqrt{2\hbar m_2\omega_{\rm m}}}
\begin{pmatrix}
m_2\omega_{\rm m} & i \\
m_2\omega_{\rm m} & -i 
\end{pmatrix}
\begin{pmatrix}
\hat{N}_X \\
\hat{P}_X
\end{pmatrix},\label{bhat}
\end{align}
where the bosonic commutation relations $[  \hat{a},\hat{a}^{\dagger}] = 1$, $[  \hat{b},\hat{b}^{\dagger}] = 1$, and $[  \hat{a},\hat{b}^{\dagger}] = 0$ etc. are assumed. 
Then, we have 
\begin{align}
\hat{H}_0
&= \hbar\omega_{\rm c}\left(  \hat{a}^{\dagger}\hat{a} + \frac{1}{2}\right)
+ \hbar\omega_{\rm m}\left(  \hat{b}^{\dagger}\hat{b} + \frac{1}{2}\right)
,\label{FQH0}\\
\hat{H}_{\rm ex}
&= i\frac{\hbar\lambda}{2} \left(  \hat{a} - \hat{a}^{\dagger}\right)\left(  \hat{b} + \hat{b}^{\dagger}\right)
+ \frac{\hbar \lambda^2}{4\omega_{\rm c}}
\left(  \hat{b} + \hat{b}^{\dagger}\right)^2
,\label{FQHex}\\
\hat{H}_{\rm dr}
&= - v_X(t)\omega_{\rm c}\sqrt{\frac{\hbar m_1\omega_{\rm c}}{2}}
\left(  \hat{a} + \hat{a}^{\dagger}\right)
.\label{FQHdr}
\end{align}
Without the driving term $\hat{H}_{\rm dr}$, the derived Hamiltonian is equivalent to a two-mode version of the Hopfield model \cite{Kockum19,Baydin25}
\begin{align}
\hat{H}
=& \hbar\omega_{\rm c}\hat{a}^{\dagger}\hat{a}
+ \hbar\omega_{\rm m}\hat{b}^{\dagger}\hat{b}
- i\hbar g\left(  \hat{a}^{\dagger}\hat{b} - \hat{a}\hat{b}^{\dagger}\right)\nonumber\\
&- i\hbar g\left(  \hat{a}^{\dagger}\hat{b}^{\dagger} - \hat{a}\hat{b}\right)+ \hbar D_{ \rm m}
\left(  \hat{b} + \hat{b}^{\dagger}\right)^2
+ {\rm const.}
,\label{Hopfield}
\end{align}
where $g = \lambda/2$ and $D_{\rm m} = \lambda^2/(4\omega_{\rm c})$. In Eq.~(\ref{Hopfield}), the term $\propto (  \hat{b} + \hat{b}^{\dagger})^2$ describes the magnon self-interaction, resulting in purely local interactions in magnetized macroscopic media. This is a magnetic analogy of the polarization self-interaction in the cavity QED, in which the electric dipole gauge provides a suitable framework for studying nonperturbative strong-coupling phenomena \cite{Todorov10,Bernardis18}.

\begin{figure}[hptb]
\begin{centering}
\includegraphics[width=0.45\textwidth,angle=0]{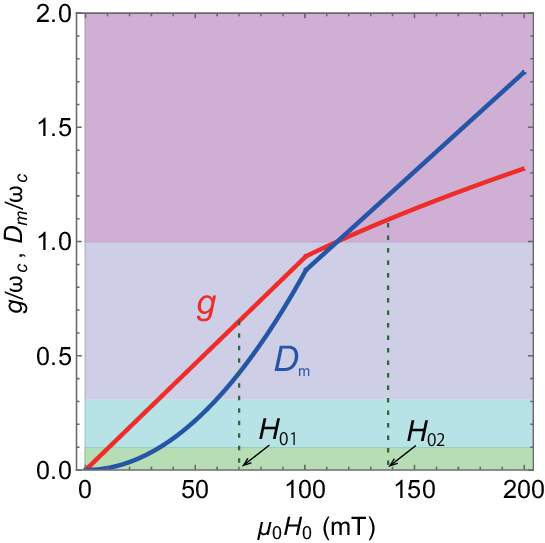}
\par\end{centering}
\caption{
Normalized coupling strength ($g/\omega_{\rm c}$) and coefficient ($D_{\rm m}/\omega_{\rm c}$) as a function of an external magnetic field ($\mu_0H_0$). For this calculation, $d_{\rm M}/d = 1$ is used.
}
\label{Fig:gDm}
\end{figure}

Assuming the cylindrical rod shape of the ferromagnet, in Fig.~\ref{Fig:gDm}, we plot the coupling strength $g$ and coefficient $D_{\rm m}$ as a function of an external magnetic field for $d_{\rm M}/d = 1$. Note that a specific value of $d_{\rm M}/d$ just determines the magnitude of these factors and does not change their lineshapes ($\mu_0H_0$-dependence). As seen, both $g$ and $D_{\rm m}$ depend on the external magnetic field $H_0$ via Eq.~(\ref{lambda}). In particular, the coupling strength $g$ coincide with Eq.~(\ref{gNRWA}) at the original modes crossing point characterized by $H_{01}$ and $H_{02}$. Also, at the critical magnetic field $\sim100$~mT, which is characterized by Eq.~(\ref{omegamrod}), a finite value of $g$ originates from the coupling between the photon mode and the soft magnon mode with $\omega_{\rm m} = 0$. 
In contrast, for a film shape of magnets, one can easily find that both $g$ and $D_{\rm m}$ become zero in the magnetic field ranges characterized by the zero-mode magnon with $\omega_{\rm m} = 0$ in Eq.~(\ref{omegamfilm}) because of $\lambda \propto \sqrt{\omega_{\parallel1}}\to0$. 

\subsection{Bloch-Siegert shift and diamagnetic-like shift} 

Here, we discuss the origin of the positive frequency shift found in Sec.~\ref{Classical model}. To this end, we redefine the derived Hopfield Hamiltonian Eq.~(\ref{Hopfield}) as the following: 
$\hat{H} = \hat{H}_{\rm RWA} + \hat{H}_{\rm CR} + \hat{H}_{\rm dia}$ with 
\begin{align}
\hat{H}_{\rm RWA}
&= \hbar\omega_{\rm c}\hat{a}^{\dagger}\hat{a}
+ \hbar\omega_{\rm m}\hat{b}^{\dagger}\hat{b}
- i\hbar g\left(  \hat{a}^{\dagger}\hat{b} - \hat{a}\hat{b}^{\dagger}\right)
,\label{QHRWA}\\
\hat{H}_{\rm CR}
&= - i\hbar g\left(  \hat{a}^{\dagger}\hat{b}^{\dagger} - \hat{a}\hat{b}\right)
,\label{QHCR}\\
\hat{H}_{\rm dia}
&= \hbar D_{ \rm m}
\left(  \hat{b} + \hat{b}^{\dagger}\right)^2
.\label{QHdia}
\end{align}
Here, $\hat{H}_{\rm RWA}$ involves the exchange-type interaction term, i.e., $\hat{a}^{\dagger}\hat{b} - \hat{a}\hat{b}^{\dagger}$, which guarantees that the excitation number of the coupled system is conserved and only states with the same number of excitations can interact. In contrast, $\hat{H}_{\rm CR}$ and $\hat{H}_{\rm dia}$ obviously involve the excitation number nonconserving terms, e.g, $\hat{a}^{\dagger}\hat{b}^{\dagger}$ and $\hat{b}^2$, leading to nontrivial frequency shifts known as the BS shift and diamagnetic-like shift, respectively \cite{Diaz19,Li18}. Accordingly, we define these two frequency shifts: the BS shift and diamagnetic-like shift by
\begin{align}
\delta\omega_\pm^{\rm BS} 
&= \omega_\pm^{\rm Dicke} - \omega_\pm^{\rm RWA}
,\label{BSshift}\\
\delta\omega_\pm^{\rm dia} 
&= \omega_\pm^{\rm Hopfield} - \omega_\pm^{\rm Dicke}
,\label{diashift}
\end{align}
where $\omega_+^{\rm Dicke} + \omega_-^{\rm Dicke} = \langle \hat{H}_{\rm RWA} + \hat{H}_{\rm CR}\rangle/(\hbar/2)$, $\omega_+^{\rm RWA} + \omega_-^{\rm RWA} = \langle \hat{H}_{\rm RWA}\rangle/(\hbar/2)$ that is equivalent to Eq.~(\ref{omegaRWA}), and $\omega_+^{\rm Hopfield} + \omega_-^{\rm Hopfield} = \langle \hat{H}\rangle/(\hbar/2)$. Here, the expectation value $\langle \cdots\rangle$ is calculated based on the ground state of the full Hamiltonian $\hat{H}$, which is obtained by $|G\rangle = \hat{U}|0\rangle$, where $|0\rangle$ is the usual vacuum state consisting of the original boson modes ($\hat{a},\hat{b}$) and $\hat{U}$ is the Hopfield-Bogolubov transformation (two-mode squeezing operator). The full Hamiltonian can be diagonalized by using $\hat{U}$ 
as 
\begin{align}
\hat{U}^{\dagger}\hat{H}\hat{U} \equiv \hat{\mathcal H} = \hbar\omega_{+}\hat{c}_{+}^{\dagger}\hat{c}_{+}
+ \hbar\omega_{-}\hat{c}_{-}^{\dagger}\hat{c}_{-}
+ {\rm const.}
,\label{Hopfield2}
\end{align}
where all dressed operators $\hat{c}_{\pm}^{(\dagger)}$ are given by the linear combination of photon ($\hat{a}^{(\dagger)}$) and magnon ($\hat{b}^{(\dagger)}$) operators \cite{Hopfield58,Ciuti05}. 
Equation~(\ref{Hopfield2}) provides the polariton eigenfrequencies:
\begin{align}
\left(  \omega_\pm^{\rm Hopfield}\right)^2
=& \frac{\omega_{\rm c}^2 + 4D_{ \rm m}\omega_{\rm m} + \omega_{\rm m}^2}{2}\nonumber\\
& \pm \sqrt{\frac{\left(  \omega_{\rm c}^2 - 4D_{ \rm m}\omega_{\rm m} - \omega_{\rm m}^2\right)^2}{4} + 4g^2\omega_{\rm c}\omega_{\rm m}}
.\label{omegaHopfield}
\end{align}
Inserting the definition of $g = \lambda/2$ and $D_{\rm m} = \lambda^2/(4\omega_{\rm c})$, the above eigenfrequencies are equivalent to Eq.~(\ref{omegaNRWA}), indicating that our circuit model of MPs corresponds to the two-mode  Hopfield model in the cavity QED. 
In addition, the quantum-classical correspondence on our models gives an important insight for the quantum phase transition (superradiant phase transition).
In usual, the superradiant phase transition occurs when the ground state of Eq.~(\ref{Hopfield2}) becomes unstable, that is, when $\omega_{-}^{\rm Hopfield}$ vanishes, which requires the condition of $D_{ \rm m} < g^2/\omega_{\rm c}$ in Eq.~(\ref{omegaHopfield}).
However, as follows from Eq.~(\ref{Hopfield}), the condition of $ D_{ \rm m} = \lambda^2/(4\omega_{\rm c}) = g^2/\omega_{\rm c}$ is always satisfied  for all parameters, indicating that the excitation energy of Eq.~(\ref{Hopfield2}) remains positive value and therefore the superradiant phase transition in the derived Hopfield model is prohibited. This implies that the vacuum state of Eq.~(\ref{Hopfield2}) is simply the ground state of the two polariton modes.
In the absence of $D_{ \rm m}$-term, the polariton  eigenfrequencies reduce to those of the two-mode Dicke model (described by $\hat{H}_{\rm RWA} + \hat{H}_{\rm CR}$)
\begin{align}
\left(  \omega_\pm^{\rm Dicke}\right)^2
= \frac{\omega_{\rm c}^2 + \omega_{\rm m}^2}{2} \pm \sqrt{\frac{\left(  \omega_{\rm c}^2 - \omega_{\rm m}^2\right)^2}{4} + 4g^2\omega_{\rm c}\omega_{\rm m}}
,\label{omegaNRWADicke}
\end{align}
which is plotted in Fig.~\ref{Fig:omega Dicke} for comparison [see Appendix~\ref{Dicke QPs}].

\begin{figure}[hptb]
\begin{centering}
\includegraphics[width=0.48\textwidth,angle=0]{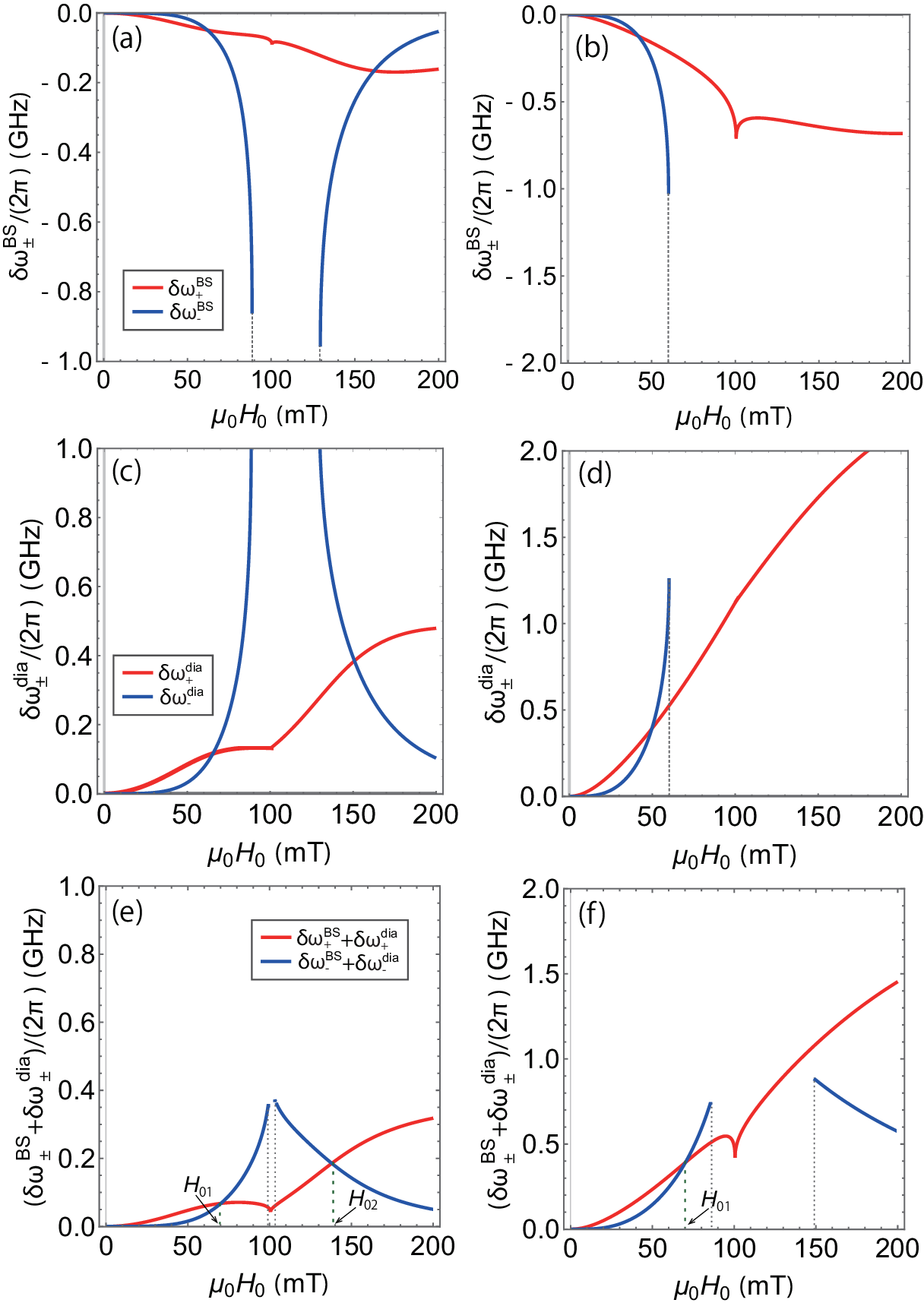}
\par\end{centering}
\caption{
(a),(b) Calculated Bloch-Siegert shift ($\delta\omega_\pm^{\rm BS}$) as a function of an external magnetic field ($\mu_0H_0$) for (a) $d_{\rm M}/d = 0.4$ and (b) $d_{\rm M}/d = 1$.
(c),(d) Calculated diamagnetic-like shift ($\delta\omega_\pm^{\rm dia}$) as a function of an external magnetic field for (a) $d_{\rm M}/d = 0.4$ and (b) $d_{\rm M}/d = 1$.
(e),(f) Sum of the two frequency shifts ($\delta\omega_\pm^{\rm BS}+\delta\omega_\pm^{\rm dia}$) as a function of an external magnetic field for (e) $d_{\rm M}/d = 0.4$ and (f) $d_{\rm M}/d = 1$.
Note that (a),(c),(e) are the nonperturbative USC case and (b),(d),(f) are the DSC case.
}
\label{Fig:BS dia shifts}
\end{figure}

\begin{figure}[hptb]
\begin{centering}
\includegraphics[width=0.45\textwidth,angle=0]{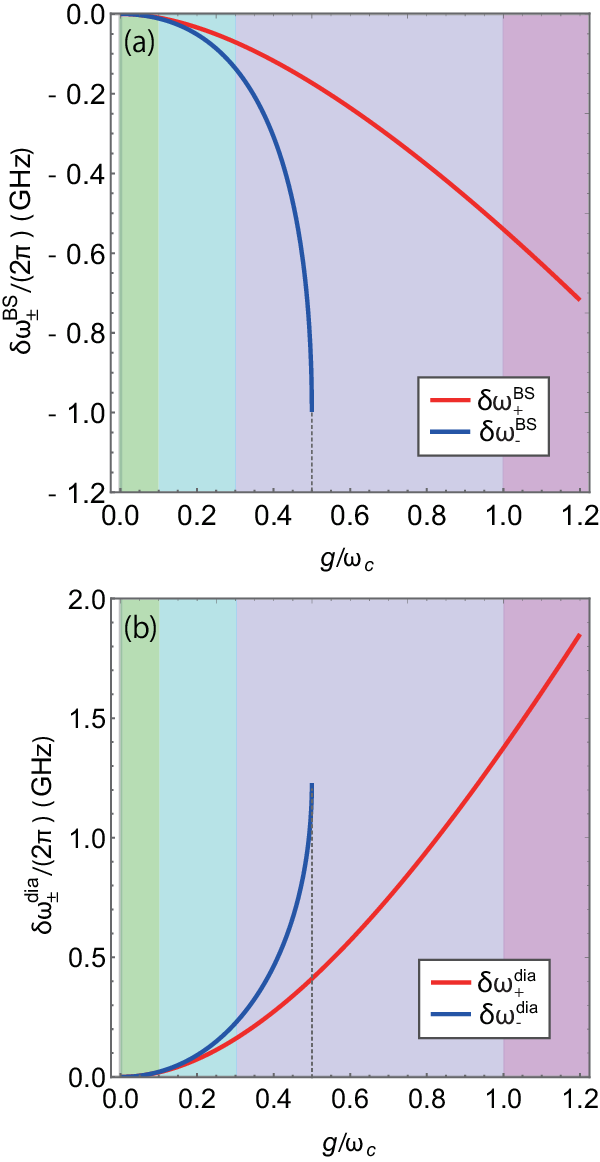}
\par\end{centering}
\caption{
(a) Calculated Bloch-Siegert shift ($\delta\omega_\pm^{\rm BS}$) at the original modes crossing point as a function of $g/\omega_{\rm c}$.
(b) Calculated diamagnetic-like shift ($\delta\omega_\pm^{\rm dia}$) at the original modes crossing point as a function of $g/\omega_{\rm c}$.
Disappearances of each shift at a certain value of $g/\omega_{\rm c}$ reflects the superradiant phase transition in the two-mode Dicke model, i.e., the breakdown of the current description.
}
\label{Fig:deltaomega,g}
\end{figure}

Figure~\ref{Fig:BS dia shifts} displays the calculated $\mu_0H_0$-dependence of the BS shift, diamagnetic-like shift, and the sum of these two shifts in the nonperturbative USC and the DSC regimes. As seen in Figs.~\ref{Fig:BS dia shifts}~(a) and (b), in the both regimes, the calculated BS shift is negative value over the displayed range of $\mu_0H_0$ as expected by the coherent magnon-photon coupling described by $\hat{H}_{\rm CR}$. Note that the disappearance of the BS shift at a certain magnetic field reflects an instability in Eq.~(\ref{omegaNRWADicke}) (superradiant phase transition in the Dicke model) due to increasing of the coherent magnon-photon coupling [see Fig.~\ref{Fig:omega Dicke} in Appendix~\ref{Dicke QPs}], i.e., the breakdown of the current description. In contrast, as shown in Figs.~\ref{Fig:BS dia shifts}~(c) and (d), the diamagnetic-like shift is positive value as expected by the magnon self-interaction described by $\hat{H}_{\rm dia}$. Similarly to the BS shift, the diamagnetic-like shift disappears at a certain magnetic field because of the phase transition occurring in the Dicke model. Now, we focus on the sum of these two shifts ($\delta\omega_\pm^{\rm BS}+\delta\omega_\pm^{\rm dia}$), which are plotted in Figs.~\ref{Fig:BS dia shifts}~(e) and (f). The resultant frequency shift is always positive value, which implies that the diamagnetic-like shift overcomes the BS shift throughout the range of $\mu_0H_0$. In fact, according to Eqs.~(\ref{BSshift}) and (\ref{diashift}), the resultant frequency shift can be written by $\delta\omega_\pm^{\rm BS}+\delta\omega_\pm^{\rm dia} = \delta\omega_\pm^{\rm Hopfield}-\delta\omega_\pm^{\rm RWA}$, which is equivalent to the frequency shift [see Eq.~(\ref{omega shift})] found in our classical circuit model. Note that the disappearance of the resultant shift at a certain magnetic field reflects the breakdown of the RWA in the nonperturbative USC and the DSC regimes. Therefore, the mechanism of the positive frequency shift in the classical circuit model can be interpreted as competition of the BS shift and the diamagnetic-like shift in the MP system.

Figure~\ref{Fig:deltaomega,g} shows the calculated BS shift and diamagnetic-like shift at the original modes crossing point, which are formulated by, respectively, 
\begin{align}
\delta\omega_\pm^{\rm BS}
&= \omega_{\rm c}\left[  \sqrt{1 \pm \frac{2g}{\omega_{\rm c}}}
- \left(  1 \pm \frac{g}{\omega_{\rm c}}\right)\right]
,\label{omegaBS shift}\\
\delta\omega_\pm^{\rm dia}
&= \omega_{\rm c}\left[  \sqrt{1 + \left(  \frac{g}{\omega_{\rm c}}\right)^2} \pm \frac{g}{\omega_{\rm c}}
- \sqrt{1 \pm \frac{2g}{\omega_{\rm c}}}\right]
,\label{omegadia shift}
\end{align}
leading to the resultant frequency $\delta\omega_\pm^{\rm BS}+\delta\omega_\pm^{\rm dia} = \delta\omega_\pm$ characterized by Eq.~(\ref{omega shift}) and Fig.~\ref{Fig:g/wc}~(b).
As seen in Fig.~\ref{Fig:deltaomega,g}, throughout the range of $\mu_0H_0$, the BS shift is negative value while the diamagnetic-like shift is positive value. Note that the disappearance of each shift at a certain value of $g/\omega_{\rm c}$ reflects the phase transition in the Dicke model. The calculated both frequency shifts start to gradually increase when the MP system goes into the perturbative USC regime. Up to the leading order of $g/\omega_{\rm c}$, we have $\delta\omega_\pm^{\rm BS} \approx -g^2/(2\omega_{\rm c})$ and $\delta\omega_\pm^{\rm dia} \approx g^2/\omega_{\rm c}$, whose tendencies are similar to that of the BS shift in typical cavity QED systems \cite{Diaz19}. 

\section{Ground-state quantum properties}\label{Quantum properties}

Based on the derived quantum mechanical model [Eq.~(\ref{Hopfield})], we calculate several quantum quantities, such as the ground-state particle number, quantum fluctuations associated with Heisenberg's principle, and entanglement entropy and connect these quantities to the positive frequency shift. We also investigate the influence of the soft magnon on these quantum quantities. 

\subsection{Average particle number} 

The presence of excitation number nonconserving terms in $\hat{H}_{\rm CR}$ and $\hat{H}_{\rm dia}$ anticipates that the ground state of Eq.~(\ref{Hopfield2}) exhibits some of nontrivial quantum properties. 
One of them for a quantity of interest is the ground-state (virtual) photon number, $\langle \hat{a}^{\dagger}\hat{a}\rangle$, which can be calculated based on the ground state $|G\rangle$ and the annihilation operator of photon in terms of the dressed ones \cite{Emary03,Ciuti05}
\begin{align}
\hat{a}
= A_{+}\hat{c}_{+} + B_{+}\hat{c}_{+}^{\dagger} + A_{-}\hat{c}_{-} + B_{-}\hat{c}_{-}^{\dagger}
,\label{}
\end{align}
where $A_+ = \cos\Theta(  \sqrt{\omega_{\rm c}/\omega_+} + \sqrt{\omega_+/\omega_{\rm c}})/2$, 
$A_- = -\sin\Theta(  \sqrt{\omega_{\rm c}/\omega_-} + \sqrt{\omega_-/\omega_{\rm c}})/2$, 
$B_+ = \cos\Theta(  \sqrt{\omega_{\rm c}/\omega_+} - \sqrt{\omega_+/\omega_{\rm c}})/2$, and
$B_- = -\sin\Theta(  \sqrt{\omega_{\rm c}/\omega_-} - \sqrt{\omega_-/\omega_{\rm c}})/2$ with $\tan2\Theta = 4g\sqrt{\omega_{\rm c}\omega_{\rm m}}/|\omega_{\rm c}^2 - 4D_{ \rm m}\omega_{\rm m} - \omega_{\rm m}^2|$.
Accordingly, the ground-state photon number is given by
\begin{align}
\left\langle \hat{a}^{\dagger}\hat{a}\right\rangle
&= B_{+}^2 + B_{-}^2\nonumber\\
&= \frac{\cos^2\Theta}{4}\frac{\left(  \omega_+ - \omega_{\rm c}\right)^2}{\omega_{\rm c}\omega_+}
+ \frac{\sin^2\Theta}{4}\frac{\left(  \omega_- - \omega_{\rm c}\right)^2}{\omega_{\rm c}\omega_-}
,\label{ground-state photon number}
\end{align}
where the expectation value $\langle \cdots\rangle$ is calculated based on $|G\rangle$ satisfied with $\hat{c}_{\pm}|G\rangle = 0$. Equation~(\ref{ground-state photon number}) implies that the vacuum state includes the higher excited states of photons due to the nonconserving terms $\hat{H}_{\rm CR}$ and $\hat{H}_{\rm dia}$. According to the similar way, one can easily calculate the ground-state magnon number as
\begin{align}
\left\langle \hat{b}^{\dagger}\hat{b}\right\rangle
&= D_{+}^2 + D_{-}^2\nonumber\\
&= \frac{\sin^2\Theta}{4}\frac{\left(  \omega_+ - \omega_{\rm m}\right)^2}{\omega_{\rm m}\omega_+}
+ \frac{\cos^2\Theta}{4}\frac{\left(  \omega_- - \omega_{\rm m}\right)^2}{\omega_{\rm m}\omega_-}
.\label{ground-state magnon number}
\end{align}
Here, the annihilation operator of magnon is expressed as 
\begin{align}
\hat{b}
= C_{+}\hat{c}_{+} + D_{+}\hat{c}_{+}^{\dagger} + C_{-}\hat{c}_{-} + D_{-}\hat{c}_{-}^{\dagger}
,\label{}
\end{align}
where $C_+ = \sin\Theta(  \sqrt{\omega_{\rm m}/\omega_+} + \sqrt{\omega_+/\omega_{\rm m}})/2$, 
$C_- = \cos\Theta(  \sqrt{\omega_{\rm m}/\omega_-} + \sqrt{\omega_-/\omega_{\rm m}})/2$, 
$D_+ = \sin\Theta(  \sqrt{\omega_{\rm m}/\omega_+} - \sqrt{\omega_+/\omega_{\rm m}})/2$, and
$D_- = \cos\Theta(  \sqrt{\omega_{\rm m}/\omega_-} - \sqrt{\omega_-/\omega_{\rm m}})/2$.

\begin{figure}[hptb]
\begin{centering}
\includegraphics[width=0.48\textwidth,angle=0]{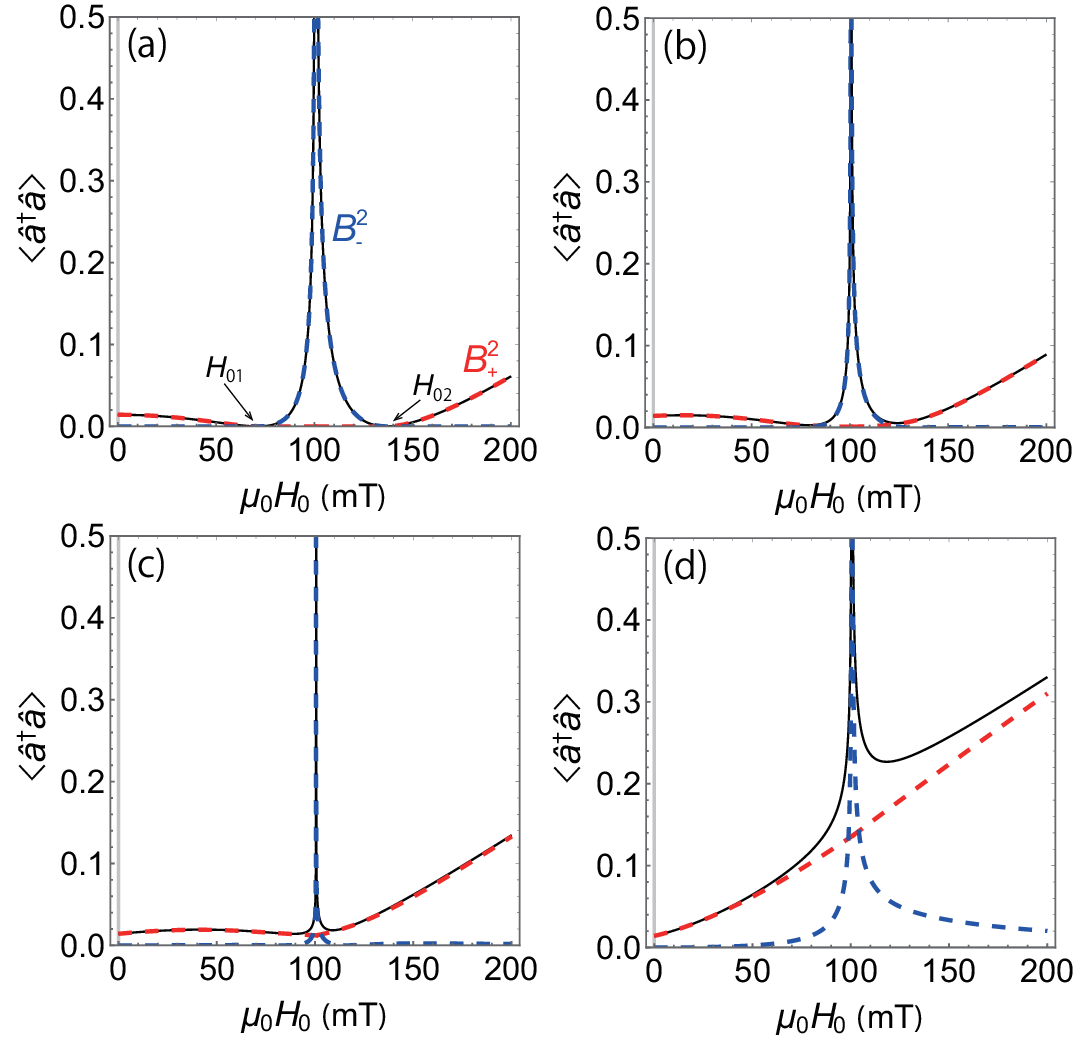}
\par\end{centering}
\caption{
Calculated ground-state photon number ($\langle \hat{a}^{\dagger}\hat{a}\rangle$) as a function of an external magnetic field ($\mu_0H_0$) for different values of $d_{\rm M}/d$: (a) 0.02, (b) 0.2, (c) 0.4, and (d) 1. The dashed red and blue lines correspond to $B_+^2$ and $B_-^2$ in Eq.~(\ref{ground-state photon number}), respectively, and the black solid line are the sum of them, i.e., $\langle \hat{a}^{\dagger}\hat{a}\rangle$. 
}
\label{Fig:ground state photon number}
\end{figure}

\begin{figure}[hptb]
\begin{centering}
\includegraphics[width=0.45\textwidth,angle=0]{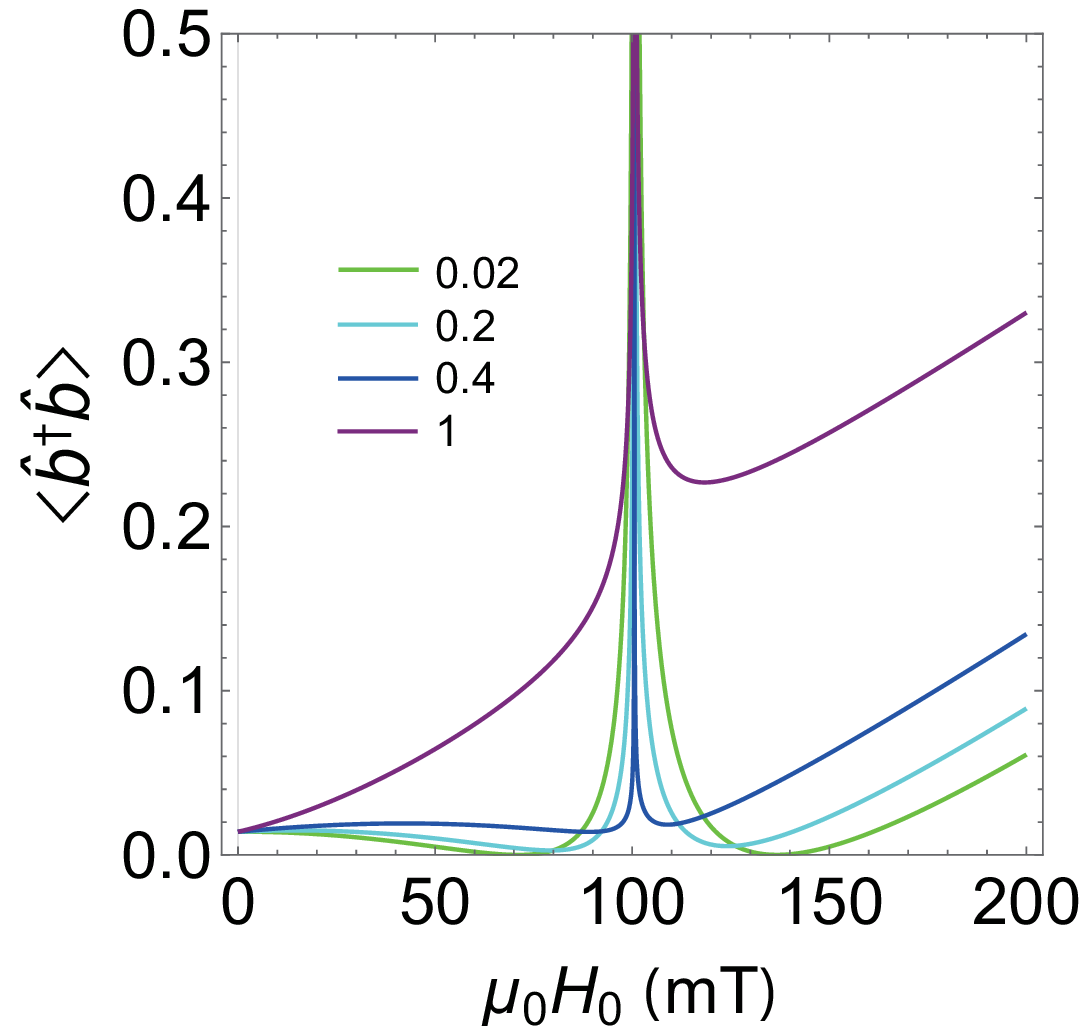}
\par\end{centering}
\caption{
Calculated ground-state magnon number ($\langle \hat{b}^{\dagger}\hat{b}\rangle$) as a function of an external magnetic field ($\mu_0H_0$) for different values of $d_{\rm M}/d = 0.02, 0.2 ,0.4 ,1$. Note that $\langle \hat{b}^{\dagger}\hat{b}\rangle = \langle \hat{a}^{\dagger}\hat{a}\rangle$.
}
\label{Fig:ground state magnon number}
\end{figure}

Figure~\ref{Fig:ground state photon number} displays the calculated ground-state photon number in different coupling strength regimes which correspond to those of Fig.~\ref{Fig:omegapm}. In the strong coupling (Fig.~\ref{Fig:omegapm}~(a)) and perturebative USC (Fig.~\ref{Fig:omegapm}~(b)) regimes, one can see $\langle \hat{a}^{\dagger}\hat{a}\rangle \approx 0$ around the original modes crossing points ($H_{01}$ and $H_{02}$). However, around the critical magnetic field characterized by Eq.~(\ref{omegamrod}), the ground-state photon number appears to diverge, $\langle \hat{a}^{\dagger}\hat{a}\rangle \to \infty$. Then, the ground state $|G\rangle$ involves the states with various excited numbers. According to Eq.~(\ref{ground-state photon number}), this divergence originates from $\omega_- = 0$ caused by the soft magnon. 
Note that recently a similar behavior of the photon (magnon) number has been reported on the context of magnonic superradiant phase transition in the cavity magnonics system \cite{Lee23}.
In the nonperturebative USC regime represented by Fig.~\ref{Fig:omegapm}~(c), not only around the critical magnetic field but also at the original modes crossing points, we can find a visible value of the ground-state photon number. 
In the DSC regime, one can see the divergence of $\langle \hat{a}^{\dagger}\hat{a}\rangle$ around the critical magnetic field as well as the rapidly grown ground-state photon number throughout the range of $\mu_0H_0$. 
Figure~\ref{Fig:ground state magnon number} shows the calculated ground-state magnon number in different coupling strength regimes which correspond to those of Fig.~\ref{Fig:ground state photon number}. Since we have the relation $\langle \hat{a}^{\dagger}\hat{a}\rangle = \langle \hat{b}^{\dagger}\hat{b}\rangle$, the $\mu_0H_0$-dependence of the ground state magnon number is equivalent to that of the ground-state photon number, which implies that the photon and magnon subsystems are correlated at the ground state $|G\rangle$. 

At the original modes crossing point, Eq.~(\ref{ground-state photon number}) can be simplified to 
\begin{align}
\left\langle \hat{a}^{\dagger}\hat{a}\right\rangle
= \frac{1}{2}\left[  \sqrt{1 + \left(  \frac{g}{\omega_{\rm c}}\right)^2} - 1\right]
.\label{ground-state photon number1}
\end{align}
Note that we have the relation $\langle \hat{a}^{\dagger}\hat{a}\rangle = \langle \hat{b}^{\dagger}\hat{b}\rangle$. For $g \ll \omega_{\rm c}$ it reduces to the from of $\langle \hat{a}^{\dagger}\hat{a}\rangle \approx (g/\omega_{\rm c})^2/4$ while for $g \gg \omega_{\rm c}$ we have $\langle \hat{a}^{\dagger}\hat{a}\rangle \approx (g/\omega_{\rm c} -1)/2$, which implies that the ground-state photon number depends on $g/\omega_{\rm c}$ alike the positive frequency shift characterized by $\delta\omega_\pm = \delta\omega_\pm^{\rm BS}+\delta\omega_\pm^{\rm dia}$ and Fig.~\ref{Fig:g/wc}~(b). In fact, it is worth noting that one can relate the ground state photon number [Eq.~(\ref{ground-state photon number})] with the positive frequency shift [Eq.~(\ref{omega shift})] by 
\begin{align}
\left\langle \hat{a}^{\dagger}\hat{a}\right\rangle
= \frac{\delta\omega_\pm}{2\omega_{\rm c}}
.\label{ground-state photon number2}
\end{align}
This relation allows to estimate the ground-state photon (magnon) number in the nonperturbative strong-coupling regimes by experimentally observing the positive frequency shift as shown in Fig.~\ref{Fig:omegapm}.

\subsection{Quantum fluctuations and entanglement entropy} 

One of the quantum quantities of interest is entanglement in the coupled system, which may be an important resource for quantum information processing.
Entanglement between the photon and magnon subsystems can be measured by the entanglement entropy $S = -{\rm Tr}\{\hat{\rho}_r\log_2\hat{\rho}_r\}$, where $\hat{\rho}_r = {\rm Tr}_{\rm magnon}|G\rangle\langle G|$ is the reduced density matrix which is obtained at zero temperature by tracing out the magnon degrees of freedom in the density matrix of the magnon-photon coupled system $\hat{\rho} = |G\rangle\langle G|$.
For a quadratic Hamiltonian of interacting bosonic fields, the entanglement entropy in terms of the Heisenberg principle is given by \cite{Nataf12}
\begin{align}
S
=& \left(  \frac{1}{\hbar}\Delta q_X\Delta\phi_X + \frac{1}{2}\right)\log_2\left(  \frac{1}{\hbar}\Delta q_X\Delta\phi_X + \frac{1}{2}\right)\nonumber\\
&- \left(  \frac{1}{\hbar}\Delta q_X\Delta\phi_X - \frac{1}{2}\right)\log_2\left(  \frac{1}{\hbar}\Delta q_X\Delta\phi_X - \frac{1}{2}\right)
,\label{entanglement entropy}
\end{align}
where $\Delta q_X = \sqrt{\langle \hat{q}_X^2\rangle - \langle \hat{q}_X\rangle^2} = \sqrt{\hbar/(2m_1\omega_{\rm c})}$$\sqrt{1 + 2\langle \hat{a}^{\dagger}\hat{a}\rangle + (\langle (\hat{a}^{\dagger})^2\rangle + \langle \hat{a}^2\rangle)}$ and $\Delta \phi_X = \sqrt{\langle \hat{\phi}_X^2\rangle - \langle \hat{\phi}_X\rangle^2} = \sqrt{\hbar m_1\omega_{\rm c}/2}$$\sqrt{1 + 2\langle \hat{a}^{\dagger}\hat{a}\rangle - (\langle (\hat{a}^{\dagger})^2\rangle + \langle \hat{a}^2\rangle)}$. 
Here, the expectation value $\langle \cdots\rangle$ is calculated based on the ground state $|G\rangle$. According to Eq.~(\ref{ahat}), the product of two quadrature variances of the original photon mode is expressed by
\begin{align} 
\Delta q_X\Delta\phi_X
=& \hbar\sqrt{\frac{1}{2} + B_{+}^2 + B_{-}^2 + \left(  A_{+}B_{+} + A_{-}B_{-}\right)}\nonumber\\
&\times \sqrt{\frac{1}{2} + B_{+}^2 + B_{-}^2 - \left(  A_{+}B_{+} + A_{-}B_{-}\right)}
,\label{Heisenberg principle}
\end{align}
where 
\begin{align} 
A_{+}B_{+} + A_{-}B_{-}
= \frac{\cos^2\Theta}{4}\frac{\omega_+^2 - \omega_{\rm c}^2}{\omega_{\rm c}\omega_+}
+ \frac{\sin^2\Theta}{4}\frac{\omega_-^2 - \omega_{\rm c}^2}{\omega_{\rm c}\omega_-}
.\label{AB}
\end{align}

\begin{figure}[hptb]
\begin{centering}
\includegraphics[width=0.45\textwidth,angle=0]{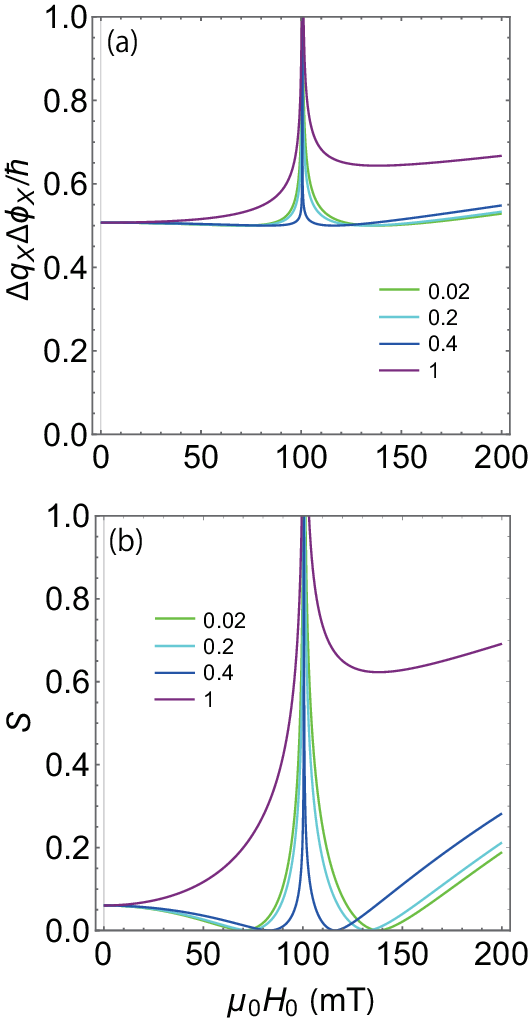}
\par\end{centering}
\caption{
Calculated (a) product of two quadrature variances of the original photon mode and (b) entanglement entropy ($S$) as a function of an external magnetic field ($\mu_0H_0$) for different values of $d_{\rm M}/d = 0.02, 0.2 ,0.4 ,1$.
}
\label{Fig:entanglement entropy}
\end{figure}

Figure~\ref{Fig:entanglement entropy}~(a) shows the calculated product of the two quadrature variances ($\Delta q_X\Delta\phi_X$) as a function of an external magnetic field. As seen, $\mu_0H_0$-dependence of $\Delta q_X\Delta\phi_X$ is very similar to that of the ground-state photon (and magnon) number. At the critical magnetic field characterized by Eq.~(\ref{omegamrod}), a quantum fluctuation between the two quadrature variances ($q_X,\phi_X$) appears to diverge, which attributes to the divergence of the ground-state photon (and magnon) number in Fig.~\ref{Fig:ground state magnon number}. Therefore, the divergence of $\Delta q_X\Delta\phi_X$ originates from the soft magnon.
Figure~\ref{Fig:entanglement entropy}~(b) shows the calculated entanglement entropy as a function of an external magnetic field. The entanglement entropy also appears to diverge at the critical magnetic field characterized by Eq.~(\ref{omegamrod}), indicating that the ground state $|G\rangle$ is strongly entangled between the photon and magnon subsystems. This is a logarithmic divergence, which attributes to 
\begin{align}
S
\approx \log_2\left(  \frac{1}{\hbar}\Delta q_X\Delta\phi_X\right)~~~({\rm for}~~~\Delta q_X\Delta\phi_X\gg\hbar/2)
.\label{entanglement entropy1}
\end{align}
Note that $\Delta q_X\Delta\phi_X\gg1$ corresponds to the divergence of the ground-state photon (and magnon) number, which means that the ground state $|G\rangle$ involves the states with various excited numbers. Such behavior of the entanglement entropy reflects a strong correlation between the ground-state photon and magnon numbers. Remarkably, at the critical magnetic field accompanied with the soft magnon, strong entanglement is obtained in not only the nonperturbative strong-coupling regime but also the usual strong coupling regime. However, in the strong coupling regime, this strong entanglement may be suppressed by the nonlinear effect of magnons such as the self-Kerr nonlinearity which is charactered by the $(\hat{b}^{\dagger}\hat{b})^2$ term \cite{Elyasi20}. 

In contrast, at the original modes crossing point, a visible entanglement entropy is obtained in only the nonperturbative strong-coupling regime. At the crossing point, Eq.~(\ref{Heisenberg principle}) can be simplified to 
\begin{align}
\Delta q_X\Delta\phi_X
= \frac{\hbar}{2}\sqrt{1 + \frac{\left(  g/\omega_{\rm c}\right)^2}{1 + \left(  g/\omega_{\rm c}\right)^2}}
= \frac{\hbar}{2}\sqrt{2 - \left(  1 + \frac{\delta\omega_\pm}{\omega_{\rm c}}\right)^{-2}}
,\label{Heisenberg principle1}
\end{align}
which enables us to experimentally evaluate the  quantum fluctuation between $q_X$ and $\phi_X$, and thereby to measure the entanglement entropy  ``in the ground state'' by measuring the positive frequency shift as shown in Fig.~\ref{Fig:omegapm}.
Considering the energy scale of microwaves with several GHz, as long as experiments are performed at room temperature, the system will be in a mixed state that contains an amount of excited states.
On the other hand, the positive frequency shift as well as the dispersion of MPs themselves can be measured without much temperature dependence, which may modify the linewidth of the transmission amplitude due to thermal fluctuations \cite{Guo23}.
Therefore, this analysis with Eq.~(\ref{Heisenberg principle1}) allows to derive the particle number fluctuations and entanglement entropy only in the ground state.
Note that the product of two quadrature variances of the original magnon mode ($\Delta N_X\Delta P_X$) satisfies the same relation of Eq.~(\ref{Heisenberg principle1}) with the upper bound $\Delta q_X\Delta\phi_X\leq (\hbar/2)\sqrt{2}$ for $g \gg \omega_{\rm c}$. Consequently, this bounding keeps the entanglement entropy finite even for $g \gg \omega_{\rm c}$, whose fact is in sharp contrast to the corresponding results of the Dicke-type Hamiltonian \cite{Nataf12} [see also Appendix~\ref{Dicke QPs}].
In addition, the two quadrature variances of the original photon mode are given by 
\begin{align}
\Delta q_X\sqrt{\frac{m_1\omega_{\rm c}}{2\hbar}}
&= \frac{1}{2}\frac{1}{\sqrt[4]{1 + \left(  g/\omega_{\rm c}\right)^2}} < \frac{1}{2}
,\\
\Delta\phi_X\sqrt{\frac{1}{2\hbar m_1\omega_{\rm c}}}
&= \frac{1}{2}\frac{\sqrt{1 + 2\left(  g/\omega_{\rm c}\right)^2}}{\sqrt[4]{1 + \left(  g/\omega_{\rm c}\right)^2}} > \frac{1}{2}
,
\end{align}
which indicates that the ground state $|G\rangle$ is intrinsically squeezed by the nonconserving terms $\hat{H}_{\rm CR}$ and $\hat{H}_{\rm dia}$ not but other driving fields.  For $g \gg \omega_{\rm c}$, the uncertainty of $q_X$ completely vanishes in one quadrature, i.e., $\Delta q_X \to 0$, anticipating that the perfect squeezing may occur in the cavity magnonics system even for the Hopfield-type model. Note that the squeezing of the original magnon mode is the same, i.e., $\Delta N_X\sqrt{m_2\omega_{\rm m}/(2\hbar)} < 1/2$ and $\Delta P_X/\sqrt{2\hbar/m_2\omega_{\rm m}} > 1/2$.

\section{Discussion and conclusion}\label{Summary}

We theoretically studied nonperturbative strong-coupling phenomena in cavity magnonics systems in which magnons in ferromagnets with various shapes interact with microwave photons of a single-mode $LC$ resonator. 
Based on an effective circuit model that accounts for the magnetic flux associated with the magnetization dynamics, we showed that nontrivial positive frequency shifts emerge in the nonperturbative strong-coupling regime characterized by $g/\omega_{\rm c}\gtrsim0.3$. 
This positive frequency shift may be experimentally observable for the fundamental mode of a chiral resonance (photon) coupled to the Kittel mode (magnon) of an anisotropic ferromagnet in magnetic metamaterials \cite{Mita25PRAP} and other cavity magnonics systems with variously shaped magnet in the nonperturbative strong coupling regime \cite{Golovchanskiy21PRAP,Ghirri23,Bourcin23}. In fact, Golovchanskiy et al. \cite{Golovchanskiy21PRAP} observed an asymmetric Rabi-like splitting with $g/\omega_{\rm c} = 0.58$ in on-chip MP systems based on superconductor/ferromagnet nanostructures. More recently, Bourcin et al. \cite{Ghirri23} reported that the standard Hopfield model including the $A^2$ term in cavity QED even fails to accurately describe MPs in the nonperturbative USC regime, in which $g/\omega_{\rm c} = 0.59$ was estimated in a ferromagnetic-slab-based MP system. By introducing a shifted magnon frequency, which becomes significant in the USC regime, they successfully reproduced the experimental MP spectra. With this feature, their magnon frequency shift may be associated with the magnon self-interaction described by Eq.~\eqref{FQHex} (or Eq.~\eqref{QHex}).

Using the circuit model, we derived a minimal quantum mechanical model for generic cavity magnonics, which includes a magnon self-interaction analogous to the $A^2$ term in the cavity QED. 
The derived quantum mechanical model is equivalent to a two-mode version of the Hopfield Hamiltonian \cite{Kockum19,Baydin25}, which explains the mechanism of the positive frequency shifts found in the {\it classical} circuit model and prevents the superradiant phase transition of the MP system even in the DSC regime.
This result stands in sharp contrast to the situation in magnon-magnon \cite{Wang24} and magnon-spin \cite{Bamba22} coupled systems, in which a minimal quantum mechanical model is given by the Dicke-type model and a magnon self-interaction is absent.
In our model, the dynamical interaction in coupled Eqs.~(\ref{MaxwellqX}) and (\ref{LLGnX}), $\lambda^2 dN_X/dt$ and $-\lambda^2 dq_X/dt$, generates not only the coherent magnon-photon coupling but also the magnon self-interaction described by Eq.~(\ref{QHex}), whose situation is similar to nonperturbative cavity QED systems in the electric dipole gauge \cite{Todorov10,Bernardis18}.

We also formulated the relation between the positive frequency shift and quantum quantities, such as the ground-state photon/magnon number, quantum fluctuations associated with Heisenberg's principle, and entanglement entropy, which provides a means to experimentally access these quantum resources relevant to quantum information processing.
As long as the experiment is performed at room temperature, the system will be in a mixed state that contains an amount of excited states while the positive frequency shift itself could be measured without much temperature dependence even under the thermal fluctuation \cite{Guo23}.
Therefore, the formulated relation enables one to nondestructively extract these quantum quantities of the ground state without measuring correlations.
Furthermore, by utilizing soft magnons in an anisotropic ferromagnet, we demonstrated that these quantum quantities diverge at a critical magnetic field at which the original magnon frequency is zero. Then, the photon and magnon subsystems are strongly correlated at the ground state that involves the states with various excited numbers. In addition, the divergence of the ground-state magnon number is associated with the enhanced magnon spin angular momentum, which may be probed by spin transport or cavity photon measurements \cite{Bauer23,Lee23}.
However, in the strong coupling regime, this divergence may be suppressed by the nonlinear effect of magnons such as the self-Kerr nonlinearity characterized by the anharmonic (biquadratic) magnon potential energy $\sim N_X^4$ \cite{Elyasi20,Tatsumi25}, where $N_X$ is the canonical position. 
In contrast, nonperturbative strong coupling overcomes the self-Kerr nonlinearity of magnon \cite{Suzuki25}, leading to the divergence-like behavior of these quantum quantities.
Our work sets the stage for cavity magnonics in the nonperturbative strong-coupling regimes by connecting the nontrivial frequency shift as an experimental observable and basic quantum quantities usually limited in the theoretical framework.

\begin{acknowledgments}
The authors thank T. Hioki, S. Yoshii, G. E. W. Bauer, and S. Tomita for valuable discussions. This work was supported by Grants-in-Aid for Scientific research (Grants No.~22K14591 and No.~25H02105). R. S. thanks to GP-Spin program at Tohoku University. H. M. acknowledges support from CSIS at Tohoku University.
\end{acknowledgments} 

\section*{Data Availability}

The data that support the findings of this article are not publicly available upon publication because it is not technically feasible and/or the cost of preparing, depositing, and hosting the data would be prohibitive within the terms of this research project. The data are available from the authors upon reasonable request.

\appendix

\section{Hamilton equations}\label{Hamilton eq}

In the classical mechanics, the Legendre transformation based on the canonical momenta Eq.~(\ref{canonical momenta}) gives the corresponding Hamiltonian: 
\begin{align}
H =& \phi_X\dot{q}_X + P_X\dot{N}_X - \mathcal{L}\nonumber\\
=& \frac{\left(  \phi_X - \lambda^2 N_X\right)^2}{2m_1} + \frac{m_1\omega_{\rm c}^2}{2}\left(  q_X - v_X(t)\right)^2\nonumber\\
&+ \frac{P_X^2}{2m_2} + \frac{m_2\omega_{\rm m}^2}{2}N_X^2
.\label{HH}
\end{align}
Note that the derived Hamiltonian appears to be the minimal coupling Hamiltonian of
electric dipoles and electromagnetic waves in the electric dipole gauge \cite{Bernardis18}.
According to the Hamilton equations, we have 
\begin{align}
\dot{q}_X
&= \frac{\partial H}{\partial \phi_X} 
= \frac{\phi_X}{m_1} - \frac{\lambda^2 N_X}{m_1}
 ,\label{qX}\\
\dot{\phi}_X
&= -\frac{\partial H}{\partial q_X}
= -m_1\omega_{\rm c}^2\left( 
 q_X - v_X(t)\right)
 ,\label{phiX}\\
\dot{N}_X
&= \frac{\partial H}{\partial p_X} 
= \frac{P_X}{m_2}
,\label{NX}\\
\dot{P}_X
&= -\frac{\partial H}{\partial N_X}
= -m_2\omega_{\rm m}^2N_X
+ \frac{\lambda^2 \phi_X}{m_1} - \frac{\lambda^4N_X}{m_1}
.\label{PX}
\end{align}
By defining a position vector on the phase space coordinates: ${\bm x}(t) = (q_X,\phi_X,N_X,P_X)$, the above equations are equivalent to
\begin{align}
\frac{d {\bm x}}{dt} = \bar{\mathcal M}{\bm x} + \bar{\mathcal F}\left(  v_X\right)
\label{H EOM}
\end{align}
where $\bar{\mathcal F}\left(  v_X\right)$ is an external driving force induced by the microwave source $V(t)$ and
\begin{align}
\bar{\mathcal M} = 
\begin{pmatrix}
0
& 1/m_1
& -\lambda^2/m_1
& 0 \\
-m_1\omega_{\rm c}^2
& 0
& 0
& 0 \\
0
& 0
& 0
& 1/m_2 \\
0
& \lambda^2/m_1
& -m_2\omega_{\rm m}^2 -\lambda^4/m_1
& 0 \\
\end{pmatrix}
.\label{H matrix}
\end{align}
Assuming a solution of Eq.~(\ref{H EOM}) as ${\bm x}(t) = {\bf u}e^{-i\omega t}$ with $\omega$ and ${\bf u}$ being an eigenfrequency and its eigenvector of $\bar{\mathcal M}$, ${\rm det}\left|  \bar{\mathcal M} + i\omega\bar{I}\right| = 0$ provides the eigenfrequency of the MP hybridized mode:
\begin{align}
\omega_\pm^2
= \frac{\omega_{\rm c}^2 + \omega_{\rm m}^2 + \lambda^2}{2} \pm \sqrt{\frac{\left(  \omega_{\rm c}^2 + \omega_{\rm m}^2 + \lambda^2\right)^2}{4} - \omega_{\rm c}^2\omega_{\rm m}^2}
.\label{omegaHPM}
\end{align}

\section{Quantum properties of the two-mode Dicke model}\label{Dicke QPs}

\begin{figure}[hptb]
\begin{centering}
\includegraphics[width=0.48\textwidth,angle=0]{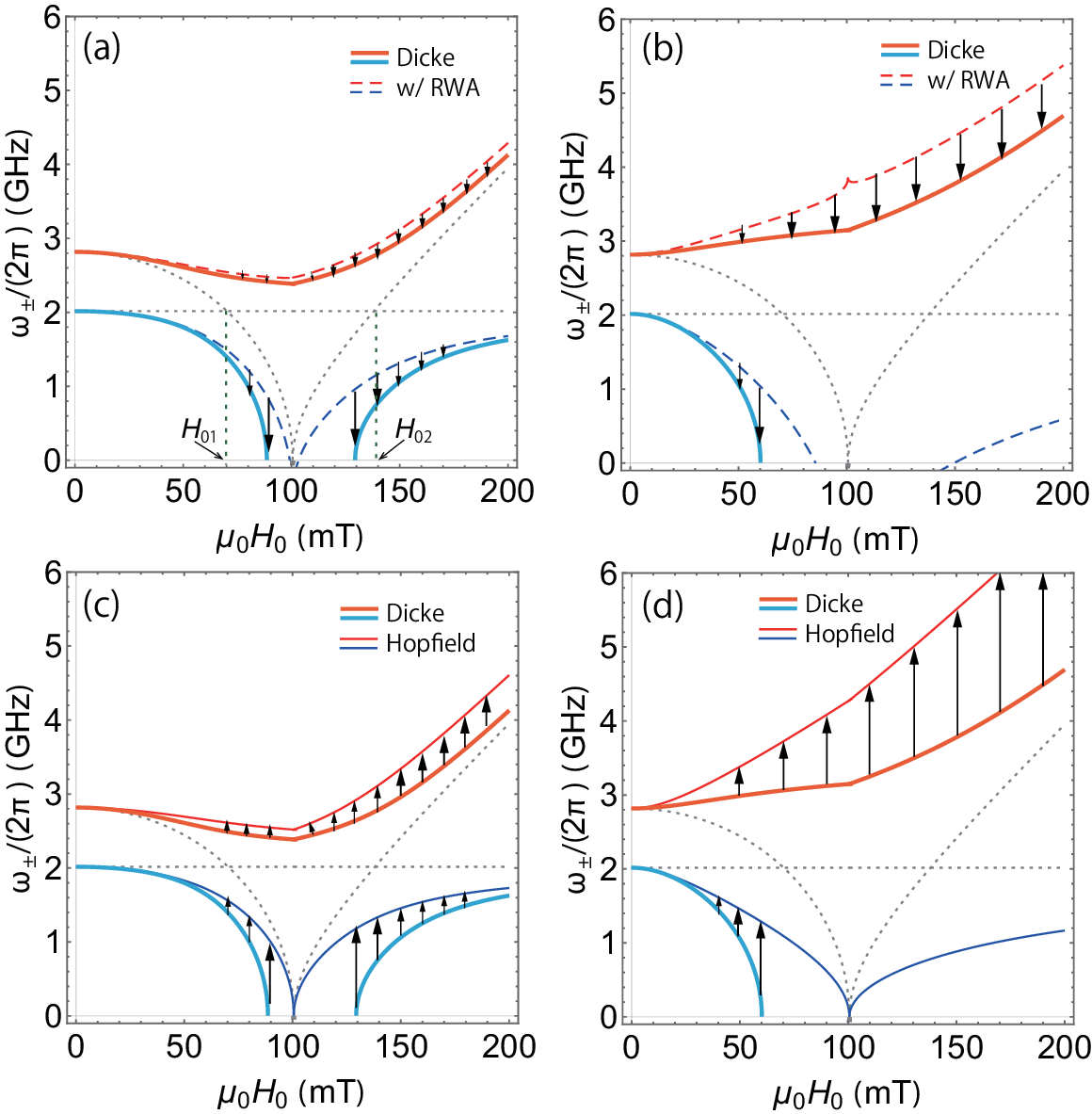}
\par\end{centering}
\caption{
(a),(b) Calculated eigenfrequency of the Dicke model in the presence of the BS shift ($\delta\omega_\pm^{\rm BS}$) as a function of an external magnetic field ($\mu_0H_0$) for (a) $d_{\rm M}/d = 0.4$ and (b) $d_{\rm M}/d = 1$.
(c),(d) Calculated eigenfrequency of the Dicke model in the presence of the diamagnetic-like shift ($\delta\omega_\pm^{\rm dia}$) as a function of an external magnetic field for (a) $d_{\rm M}/d = 0.4$ and (b) $d_{\rm M}/d = 1$.
Each black arrow represents the BS ($\downarrow$) or diamagnetic-like ($\uparrow$) shifts at each external magnetic field. Note that (a),(c) are the nonperturbative USC case and (b),(d) are the DSC case.
}
\label{Fig:omega Dicke}
\end{figure}

While the derived quantum mechanical model in Sec.~\ref{Quantum model} is identified with the two-mode version of the Hopfield Hamiltonian, it is worth considering the $D_{\rm m} = 0$ case, namely, the corresponding Dicke-type Hamiltonian. Figure~\ref{Fig:omega Dicke} shows the calculated eigenfrequency of the two-mode Dicke model ($\omega_{\pm}^{\rm Dicke}$ in Eq.~(\ref{omegaNRWADicke})) for different values of $d_{\rm M}/d$. As seen, the lower polariton branch partially disappears at a certain magnetic field, which indicates an instability in Eq.~(\ref{omegaNRWADicke}) (superradiant phase transition in the Dicke model) in the nonperturbative USC and the DSC regimes.

Regarding the original modes crossing point, we calculate the ground-state photon/magnon number, a quantum fluctuation between the two quadrature variances ($q_X,\phi_X$), and each quadrature variance for Eq.~(\ref{Hopfield}) by setting $D_{\rm m} = 0$. Then, the ground-state photon/magnon number is 
\begin{align}
&\left\langle \hat{a}^{\dagger}\hat{a}\right\rangle
= \left\langle \hat{b}^{\dagger}\hat{b}\right\rangle\nonumber\\
&= \frac{1}{2}\left[  \frac{\left(  \sqrt{1 + \frac{2g}{\omega_{\rm c}}} + \sqrt{1 - \frac{2g}{\omega_{\rm c}}}\right)
\left(  \sqrt{1 - 4\left(  \frac{g}{\omega_{\rm c}}\right)^2} + 1\right)}{4\sqrt{1 - 4\left(  \frac{g}{\omega_{\rm c}}\right)^2}} - 1\right]
,\label{ground-state photon number1D}
\end{align}
which is also expressed as
\begin{align}
\left\langle \hat{a}^{\dagger}\hat{a}\right\rangle
= \left\langle \hat{b}^{\dagger}\hat{b}\right\rangle
= \frac{1}{2}\left(  \sinh^2r_{+} + \sinh^2r_{-}\right)
,\label{ground-state photon number1D}
\end{align}
where $r_\pm = \log\left(  1 \pm 2g/\omega_{\rm c}\right)^{1/4}$ are the squeezing parameters.
According to Eq.~(\ref{Heisenberg principle}), the product of two quadrature variances of the original photon mode is 
\begin{align}
\Delta q_X\Delta\phi_X
= \frac{\hbar}{2}\frac{\sqrt{1 + 2g/\omega_{\rm c}} + \sqrt{1 - 2g/\omega_{\rm c}}}{2\sqrt[4]{1 - 4\left(  g/\omega_{\rm c}\right)^2}}
.\label{Heisenberg principle1D}
\end{align}
Note that the product of two quadrature variances of the original magnon mode ($\Delta N_X\Delta P_X$) satisfies the same relation of Eq.~(\ref{Heisenberg principle1D}) and diverges at the critical point $g = \omega_{\rm c}/2$ \cite{Nataf12}. The corresponding entanglement entropy is also, that is, $S\approx \log_2\left(  \Delta q_X\Delta\phi_X/\hbar\right) \to \infty$ at the critical point even for the original modes crossing point. These divergences are sharply in contrast to the corresponding results of the Hopfield Hamiltonian in Sec.~\ref{Quantum properties}. On the other hand, the two quadrature variances of the original photon mode are given by
\begin{align}
\Delta q_X\sqrt{\frac{m_1\omega_{\rm c}}{2\hbar}}
&= \frac{1}{2}\sqrt{\frac{\sqrt{1 + 2g/\omega_{\rm c}}}{2} + \frac{\sqrt{1 - 2g/\omega_{\rm c}}}{2}} < \frac{1}{2}
,\\
\Delta\phi_X\sqrt{\frac{1}{2\hbar m_1\omega_{\rm c}}}
&= \frac{1}{2}\sqrt{\frac{1}{2\sqrt{1 + 2g/\omega_{\rm c}}} + \frac{1}{2\sqrt{1 - 2g/\omega_{\rm c}}}} > \frac{1}{2}
,
\end{align}
indicating that the ground state $|G\rangle$ is intrinsically squeezed by the nonconserving term $\hat{H}_{\rm CR}$ for $g < \omega_{\rm c}/2$ \cite{Emary03}. Note that the squeezing of the original magnon mode ($\hat{b}$) is the same.



\begin{thebibliography}{999}

\bibitem{Walther06}H. Walther, B. T. H. Varcoe, B.-G. Englert, and T. Becker, Cavity quantum electrodynamics, Rep. Prog. Phys. {\bf 69}, 1325 (2006).


\bibitem{Blais21}A. Blais, A. L. Grimsmo, S. M. Girvin, and A. Wallraff, Circuit quantum electrodynamics, Rev. Mod. Phys. {\bf 93}, 025005 (2021).

\bibitem{Diaz19}P. F.-D\'{i}az, L. Lamata, E. Rico, J. Kono, and E. Solano, Ultrastrong coupling regimes of light-matter interaction, Rev. Mod. Phys. {\bf 91} 025005 (2019).

\bibitem{Kockum19}A. F. Kockum, A. Miranowicz, S. D. Liberato, S. Savasta, and F. Nori, Ultrastrong coupling between light and matter, Nat. Rev. Phys. {\bf 1}, 19 (2019).

\bibitem{Schlawin22}F. Schlawin, D. M. Kennes, and M. A. Sentef, Cavity quantum materials, Appl. Phys. Rev. {\bf 9}, 011312 (2022).

\bibitem{Baydin25}A. Baydin, H. Zhu, M. Bamba, K. R. A. Hazzard, and J. Kono, Perspective on the quantum vacuum in matter, Opt. Mater. Express {\bf 8}, 1833 (2025).

\bibitem{Diaz10}P. F.-D\'{i}az, J. Lisenfeld, D. Marcos, J. J. Garc\'{i}a-Ripoll, E. Solano, C. J. P. M. Harmans, and J. E. Mooij, Observation of the Bloch-Siegert shift in a qubit-oscillator system in the ultrastrong coupling regime, Phys. Rev. Lett. {\bf 105}, 237001 (2010).

\bibitem{Li18}X. Li, M. Bamba, Q. Zhang, S. Fallahi, G. C. Gardner, W. Gao, M. Lou, K. Yoshioka, M. J. Manfra, and J. Kono, Vacuum Bloch-Siegert Shift in Landau Polaritons with Ultra-high Cooperativity, Nat. Photonics {\bf 12}, 324 (2018).

\bibitem{Hirokawa17}M. Hirokawa, J. S. Møller, and I. Sasaki, A mathematical analysis of dressed photon in ground state of generalized quantum Rabi model using pair theory, J. Phys. A: Math. Theo. {\bf 50}, 184003
(2017).

\bibitem{Yoshihara18}F. Yoshihara, T. Fuse, Z. Ao, S. Ashhab, K. Kakuyanagi, S. Saito, T. Aoki, K. Koshino, and K. Semba, Inversion of qubit energy levels in qubit-oscillator circuits in the deepstrong-coupling regime, Phys. Rev. Lett. {\bf 120}, 183601 (2018).

\bibitem{Hepp73}K. Hepp and E. H. Lieb, On the superradiant phase transition for molecules in a quantized radiation field: The Dicke maser model, Ann. Phys. {\bf 76}, 360 (1973).

\bibitem{Bernardis18}D. De Bernardis, T. Jaako, and P. Rabl, Cavity quantum electrodynamics in the nonperturbative regime, Phys. Rev. A {\bf 97}, 043820 (2018).

\bibitem{Hayashida23}K. Hayashida, T. Makihara, N. Marquez Peraca, D. Fallas Padilla, H. Pu, J. Kono, and M. Bamba, Perfect intrinsic squeezing at the superradiant phase transition critical point, Sci. Rep. {\bf 13}, 2526 (2023).

\bibitem{Ferraro18}D. Ferraro, M. Campisi, G. M. Andolina, V. Pellegrini, and M. Polini, High-Power Collective Charging of a Solid-State Quantum Battery, Phys. Rev. Lett. {\bf 120}, 117702 (2018).

\bibitem{Chumak22}A. V. Chumak, P. Kabos, M. Wu, C. Abert, C. Adelmann {\it et al.}, Advances in magnetics: roadmap on spin-wave computing, IEEE Trans. Magn. {\bf 58}, 1 (2022).

\bibitem{Harder21}M. Harder, B. M. Yao, Y. S. Gui, C.-M. Hu, Coherent and dissipative cavity magnonics, J. Appl. Phys. \textbf{129}, 201101 (2021).

\bibitem{Rameshti22}B. Z. Rameshti, S. V. Kusminskiy, J. A. Haigh, K. Usami, D. L.-Quirion, Y. Nakamura, C.-M. Hu, H. X. Tang, G. E. Bauer, and Y. M. Blanter, Cavity magnonics, Phys. Rep. {\bf 979}, 1 (2022).

\bibitem{Yuan22}H.Y. Yuan, Y. Cao, A. Kamra, R. A. Duine, P. Yan, Quantum magnonics: when magnon spintronics meets quantum information science, Phys. Rep. {\bf 965}, 26 (2022).

\bibitem{Elyasi20}M. Elyasi, Y. M. Blanter, and G. E. W. Bauer, Resources of nonlinear cavity magnonics for quantum information, Phys. Rev. B {\bf 101}, 054402 (2020).

\bibitem{Yang21}Z.-B. Yang, H. Jin, J.-W. Jin, J.-Y. Liu, H.-Y. Liu, and R.-C. Yang, Bistability of squeezing and entanglement in cavity magnonics, Phys. Rev. Research {\bf 3}, 023126 (2021).

\bibitem{Lee23}J. M. Lee, H.-W. Lee, and M.-J. Hwang, Cavity magnonics with easy-axis ferromagnets: Critically enhanced magnon squeezing and light-matter interaction, Phys. Rev. B {\bf 108}, L241404 (2023).

\bibitem{Kani25}A. Kani, M. Hatifi, and J. Twamley, Squeezed microwave and magnonic frequency combs, APL Quantum {\bf 2}, 016112 (2025).

\bibitem{Wan24}Q.-K. Wan, H.-L. Shi, and X.-W. Guan, Quantum-enhanced metrology in cavity magnonics, Phys. Rev. B {\bf 109}, L041301 (2024).

\bibitem{Sun21}F.-X. Sun, S.-S. Zheng, Y. Xiao, Q. Gong, Q. He, and K. Xia, Remote generation of magnon schr\''{o}dinger cat state via magnon-photon entanglement, Phys. Rev. Lett. {\bf 127}, 087203 (2021).

\bibitem{Kounalakis22}M. Kounalakis, G. E. W. Bauer, and Y. M. Blanter, Analog quantum control of magnonic cat states on a chip by a superconducting qubit, Phys. Rev. Lett. {\bf 129}, 037205 (2022).

\bibitem{Mousolou21}V. A. Mousolou, Y. Liu, A. Bergman, A. Delin, O. Eriksson, M. Pereiro, D. Thonig, and E. Sj\''{o}qvist, Magnon-magnon entanglement and its quantification via a microwave cavity, Phys. Rev. B {\bf 104}, 224302(R) (2021).

\bibitem{Silaev23}M. Silaev, Ultrastrong magnon-photon coupling, squeezed vacuum, and entanglement
in superconductor/ferromagnet nanostructures, Phys. Rev. B {\bf 107}, L180503 (2023).

\bibitem{Yao23}B. Yao, Y. S. Gui, J. W. Rao, Y. H. Zhang, W. Lu, and C.-M. Hu, Coherent microwave emission of gain-driven polaritons, Phys. Rev. Lett. {\bf 130}, 146702 (2023).

\bibitem{Zhang25}C. Zhang, M. Kim, Y.-H. Zhang, Y.-P. Wang, D. Trivedi, A. Krasnok, J. Wang, D. Isleifson, R. Roshko, and C.-M. Hu, Gainloss coupled systems, APL Quantum {\bf 2}, 011501 (2025).

\bibitem{Kamra16}A. Kamra and W. Belzig, Super-Poissonian Shot Noise of Squeezed-Magnon Mediated Spin Transport, Phys. Rev. Lett. {\bf 116}, 146601 (2016).

\bibitem{Golovchanskiy21SciAdv}I. A. Golovchanskiy, N. N. Abramov, V. S. Stolyarov, M. Weides, V. V. Ryazanov, A. A. Golubov, A. V. Ustinov, and M. Y. Kupriyanov, Ultrastrong photon-to-magnon coupling in multilayered heterostructures involving superconducting coherence via ferromagnetic layers, Sci. Adv. {\bf 7}, eabe8638 (2021).

\bibitem{Golovchanskiy21PRAP}I. Golovchanskiy, N. N. Abramov, V. S. Stolyarov, A. A. Golubov, M. Yu. Kupriyanov, V. V. Ryazanov, and A. V. Ustinov, Approaching deep-strong on-chip photon-tomagnon coupling, Phys. Rev. Appl. {\bf 16}, 034029 (2021).

\bibitem{Ghirri23}A. Ghirri, C. Bonizzoni, M. Maksutoglu, A. Mercurio, O. Di Stefano, S. Savasta, and M. Affronte, Ultrastrong magnon-photon coupling achieved by magnetic films in contact with superconducting resonators, Phys. Rev. Appl. {\bf 20}, 024039 (2023).

\bibitem{Bourcin23}G. Bourcin, J. Bourhill, V. Vlaminck, and V. Castel, Strong to ultrastrong coherent coupling measurements in a YIG/cavity system at room temperature, Phys. Rev. B {\bf 107}, 214423 (2023).

\bibitem{Mita25PRAP}K. Mita, T. Chiba, T. Kodama, T. Ueda, T. Nakanishi, K. Sawada, S. Tomita, Ultrastrongly coupled and directionally nonreciprocal magnon polaritons in magnetochiral metamolecules, Phys. Rev. Appl. {\bf 23}, L011004 (2025).

\bibitem{Mita25arXiv}K. Mita, T. Kodama, T. Nakanishi, T. Ueda, K. Sawada, T. Chiba, S. Tomita, Microwave One-way Transparency by Large Synthetic Motion of Magnetochiral Polaritons in Metamolecules,	arXiv:2503.22279.

\bibitem{Yuan21}H. Y. Yuan and Rembert A. Duine, Universal field dependence of magnetic resonance near zero frequency, Phys. Rev. B {\bf 103}, 134440 (2021).

\bibitem{Bauer23}G. E. W. Bauer, P. Tang, M. Elyasi, Y. M. Blanter, and B. J. van Wees, Soft magnons in anisotropic ferromagnets, Phys. Rev. B {\bf 108}, 064431 (2023).

\bibitem{Iihama14}S. Iihama, S. Mizukami, H. Naganuma, M. Oogane, Y. Ando, and T. Miyazaki, Gilbert damping constants of Ta/CoFeB/MgO(Ta) thin films measured by optical detection of precessional magnetization dynamics, Phys. Rev. B {\bf 89}, 174416 (2014).

\bibitem{de Wal23}D. K. de Wal, A. Iwens, T. Liu, P. Tang, G. E. W. Bauer, and B. J. van Wees, Long distance magnon transport in the van der Waals antiferromagnet CrPS4, Phys. Rev. B {\bf 107}, L180403 (2023).

\bibitem{Bai15}L. Bai, M. Harder, Y. P. Chen, X. Fan, J. Q. Xiao, and C.-M. Hu, Spin Pumping in Electrodynamically Coupled Magnon-Photon Systems, Phys. Rev. Lett. {\bf 114}, 227201 (2015).

\bibitem{Grigoryan18}V. L. Grigoryan, K. Shen, and K. Xia, Synchronized spin-photon coupling in a microwave cavity, Phys. Rev. B {\bf 98}, 024406 (2018).

\bibitem{Chiba24APL}T. Chiba, T. Komine, and T. Aono, Ultrastrong-coupled magnon-polariton in a dynamical inductor based on magnetic-insulator/topological-insulator bilayers, Appl. Phys. Lett. {\bf 124}, 012402 (2024).

\bibitem{Chiba24MSJ}T. Chiba, T. Komine, and T. Aono, Microwave Transmission Theory for On-Chip Ultrastrong-Coupled Magnon-Polariton in Dynamical Inductors, J. Mag. Soc. Jpn. {\bf 48}, 21 (2024).

\bibitem{Hoffman13}S. Hoffman, K. Sato, and Y. Tserkovnyak, Landau-Lifshitz theory of the longitudinal spin Seebeck effect, Phys. Rev. B {\bf 88}, 064408 (2013).

\bibitem{Cornelissen16}L. J. Cornelissen, K. J. H. Peters, G. E. W. Bauer, R. A. Duine, and B. J. van Wees, Magnon Spin Transport Driven by the Magnon Chemical Potential in a Magnetic Insulator, Phys. Rev. B {\bf 94}, 014412 (2016).

\bibitem{Todorov10}Y. Todorov, A. M. Andrews, R. Colombelli, S. De Liberato, C. Ciuti, P. Klang, G. Strasser, and C. Sirtori, Ultrastrong lightmatter coupling regime with polariton dots, Phys. Rev. Lett. {\bf 105}, 196402 (2010).

\bibitem{Hopfield58}J. J. Hopfield, Theory of the Contribution of Excitons to the Complex Dielectric Constant of Crystals, Phys. Rev. {\bf 112}, 1555 (1958).

\bibitem{Ciuti05}C. Ciuti, G. Bastard, and I. Carusotto, Quantum vacuum properties of the intersubband cavity polariton field, Phys. Rev. B {\bf 72}, 115303 (2005).

\bibitem{Emary03}C. Emary and T. Brandes, Chaos and the quantum phase transition in the Dicke model, Phys. Rev. E {\bf 67}, 066203 (2003).

\bibitem{Nataf12}P. Nataf, M. Dogan, and K. L. Hur, Heisenberg uncertainty principle as a probe of entanglement entropy: Application to superradiant quantum phase transitions, Phys. Rev. A {\bf 86}, 043807 (2012).

\bibitem{Guo23}S. Guo, D. Russell, J. Lanier, H. Da, P. C. Hammel, and F. Yang, Strong on-chip microwave photon-magnon coupling using ultralow-damping epitaxial ${\rm Y_3Fe_5O_{12}}$ films at 2 K, Nano Lett. {\bf 23}, 5055 (2023). 

\bibitem{Wang24}Y. Wang, Y. Zhang, C. Li, J. Wei, B. He, H. Xu, J. Xia, X. Luo, J. Li, J. Dong, et al., Ultrastrong to nearly deep-strong magnon-magnon coupling with a high degree of freedom in synthetic antiferromagnets, Nat. Commun. {\bf 15}, 2077 (2024).

\bibitem{Bamba22}M. Bamba, X. Li, N. Marquez Peraca, and J. Kono, Magnonic superradiant phase transition, Commun. Phys. {\bf 5}, 3 (2022).

\bibitem{Tatsumi25}R. Tatsumi, T. Chiba, T. Komine, and H. Matsueda, Chaotic magnetization dynamics in magnetic Duffing oscillator, Phys. Rev. E {\bf 111}, 064202 (2025).

\bibitem{Suzuki25}R. Suzuki, T. Chiba, and H. Matsueda, Gain-driven magnon-polariton dynamics in the ultrastrong coupling regime: 
Effective circuit approach for coherence versus nonlinearity, arXiv:2509.09117.

\end{thebibliography}

\end{document}